\documentclass[sigconf,edbt]{acmart-edbt2020}

\def\BibTeX{{\rm B\kern-.05em{\sc i\kern-.025em b}\kern-.08em
    T\kern-.1667em\lower.7ex\hbox{E}\kern-.125emX}}

\usepackage{booktabs} 
\usepackage{graphicx}
\usepackage{subfig}
\usepackage{booktabs}
\usepackage{wrapfig}
\usepackage{enumitem}
\usepackage{url}
\usepackage{hyperref}
\usepackage{todonotes}
\usepackage{algorithm}
\usepackage[noend]{algpseudocode}

\usepackage{tikz}

\usepackage[normalem]{ulem}

\def\BState{\State\hskip-\ALG@thistlm}

\algnewcommand\algorithmicswitch{\textbf{switch}}
\algnewcommand\algorithmiccase{\textbf{case}}
\algnewcommand\algorithmicforeach{\textbf{for each}}
\algnewcommand{\IfThenElse}[3]{
	\State \algorithmicif\ #1\  \algorithmicthen\ \State #2\ \State \algorithmicelse\ \State #3}
\algdef{SE}[SWITCH]{Switch}{EndSwitch}[1]{\algorithmicswitch\ #1\ \algorithmicdo}{\algorithmicend\ \algorithmicswitch}%
\algdef{SE}[CASE]{Case}{EndCase}[1]{\algorithmiccase\ #1}{\algorithmicend\ \algorithmiccase}%
\algtext*{EndSwitch}%
\algtext*{EndCase}%
\algdef{S}[FOR]{ForEach}[1]{\algorithmicforeach\ #1\ \algorithmicdo}
\algdef{SE}[DOWHILE]{Do}{doWhile}{\algorithmicdo}[1]{\algorithmicwhile\ #1}
\algdef{S}[IF]{LIf}[2]{\parbox[t]{\dimexpr\linewidth- #1}{\algorithmicif \ #2\ \algorithmicthen \strut}}
{\algtext*{EndOn}}

\setcopyright{rightsretained}

\begin{document}
\title{MV-PBT: Multi-Version Index for Large Datasets and HTAP Workloads}
  
\author{Christian Riegger}
\affiliation{%
  \institution{Data Management Lab, \\ Reutlingen University, Germany}
 }
 \email{christian.riegger@reutlingen-university.de}

\author{Tobias Vin\c{c}on}
\affiliation{%
  \institution{Data Management Lab, \\ Reutlingen University, Germany}
 }
\email{tobias.vincon@reutlingen-university.de}

\author{Robert Gottstein}
\affiliation{%
  \institution{Data Management Lab, \\ Reutlingen University, Germany}
 }
\email{robert.gottstein@reutlingen-university.de}

\author{Ilia Petrov}
\affiliation{%
  \institution{Data Management Lab, \\ Reutlingen University, Germany}
 }
\email{ilia.petrov@reutlingen-university.de}

\renewcommand{\shortauthors}{}

\begin{abstract}
Modern mixed (HTAP) workloads execute fast update-transactions and long-running analytical queries on the same dataset and system. In multi-version (MVCC) systems, such workloads result in many short-lived versions and long version-chains as well as in increased and frequent maintenance overhead. 

Consequently, the \emph{index pressure} increases significantly. Firstly, the frequent modifications cause frequent creation of new versions, yielding a surge in index maintenance overhead. Secondly and more importantly, index-scans incur extra I/O overhead to determine, which of the resulting tuple-versions are visible to the executing transaction (visibility-check) as current designs only store version/timestamp information in the base table -- not in the index. Such \emph{index-only visibility-check} is critical for HTAP workloads on large datasets.

In this paper we propose the \emph{Multi-Version Partitioned B-Tree (MV-PBT)} as a version-aware index structure, supporting index-only visibility checks and flash-friendly I/O patterns. The experimental evaluation indicates a 2x improvement for analytical queries and 15\% higher transactional throughput  under HTAP workloads (CH-Benchmark). MV-PBT offers 40\% higher transactional throughput compared to WiredTiger's LSM-Tree implementation under YCSB.

\end{abstract}

%
%



\maketitle

\section{Introduction}
\label{sec:intro}
The spread of large-scale, data-intensive, real-time analytical applications is increasing. Such applications result in Hybrid Transactional and Analytical Processing workloads (\emph{HTAP}) combining long running analytical queries (OLAP) as well as frequent and low-latency update transactions (OLTP) on the same dataset and even on the same system \cite{Ozcan:HTAP:SIGMOD:2017}.

Multi-versioning is at the core of many approaches and system designs suitable for HTAP. Under \emph{Multi-Version Concurrency Control (MVCC)} reading transactions, executing long-running queries, do not block the frequent low-latency modifying transactions. Under such approaches multiple versions of each data item (i.e. tuple) may physically co-exist, whereas every transaction operates against a snapshot of the database comprising all versions it is allowed to see for consistent execution. Read operations simply operate on the latest  committed version, visible to them and are therefore never blocked, yielding good read performance and concurrency. An update operation produces a new version of the updated data item and invalidates the predecessor version. All versions of tuple form a version-chain. Timestamps placed on every physical version-record are used to determine, which of the exisiting tuple-versions is \emph{visible} to a transaction.

\begin{figure}
	\centering
		\hspace{-2pt}
		\includegraphics[scale=1.35]{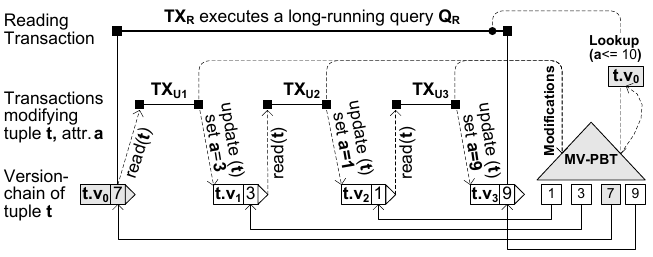}
		\vspace{-15pt}
	\caption{HTAP and Version-Chain Lengths:
	$TX_{U1}$ \dots $TX_{U3}$ create new versions of tuple \emph{t}, which are indexed. The index scan of $TX_{R}$ returns only the index entries ($t.v_{0}$) visible to $TX_{R}$ filtering the invisible ones ($t.v_{1}$ \dots $t.v_{3}$), matching the search predicate.}
	\label{fig:headpcit}
	\vspace{-15pt}
\end{figure}

Under OLTP workloads, version-chains tend to be short, due to the predominantly short-lived transactions. For instance, under TPC-C the average version-chain length is approx. 1.2 \cite{Gottstein:Diss:2016}. \emph{Under HTAP the DBMS needs to handle much longer version-chains due to the mix of long-running and short-lived transactions (Figure \ref{fig:headpcit}).} Whenever a transaction $TX_{R}$ reads a tuple $t$ the DBMS returns the latest version of that tuple $t.v_{0}$, committed before the start of $TX_{R}$. Even though, in the meantime multiple low-latency updating transactions $TX_{U1}$ \dots $TX_{U3}$ might have committed, producing successor-versions ($t.v_{1}$ \dots $t.v_{1}$. $t.v_{3}$), $t.v_{0}$ cannot be garbage collected as long as, it is visible to an active transaction, i.e. $TX_{R}$. Thus, the amount of such transient versions can be as high as \emph{several hundred millions} in real systems \cite{Lee:HybridMVCC:GC:SIGMOD:2016}. 

\emph{HTAP workloads in combination with long version-chains exercise significant pressure on indices.} In a single-versioned system there is one index entry per tuple. However, in a multi-versioned system, the DBMS needs to index at least all committed tuple-versions (Figure \ref{fig:headpcit}), even the transient ones. Given long version-chains thus put extra pressure on the index. \emph{Although most of today's systems are multi-versioned, the majority of index approaches still handle tuple-versions of the same tuple as if they were separate tuples, ignoring the version semantics.} 
If na\"ively integrated, these slow down index lookups and may cause significant maintenance overhead to persistent indices, as index updates are very frequent and since index entries corresponding to obsolete tuple-versions need to be frequently garbage collected.  Given the read/write asymmetry of modern persistent storage technologies these operations result in prohibitively expensive in-place updates. In this context append-based index structures trading sequential writes for complex reads are a good candidate.

All in all, the following observations can be made:\\
1) \emph{Version-obliviousness:} Although, all tuple-versions need to be indexed, current indexing approaches lack version information.\\
2) \emph{No index-only visibility-check:} Currently, it is impossible to determine which  index-entries resulting from an index lookup/scan correspond to versions, visible to the calling transaction.\\
3) \emph{I/O overhead:} Version-oblivious indices or na\"ive support for multi-versioning yield signifiant I/O overhead.

In the present paper we propose \emph{Multi-Version Partitioned B-Trees (MV-PBT)} as a version-aware index structure for MV-DBMS, in an attempt to address the above issues. MV-PBT is based on a variant of B\textsuperscript{+}-Trees called \emph{Partitioned B-Trees} \cite{graefe:sortingindexingpbt}. The contributions of this paper are:
\begin{itemize}[leftmargin=*,noitemsep,nolistsep]
\item  MV-PBT is a version-aware index structure. It contains version information and supports index-only visibility-checks. 
	
\item MV-PBT support append-based write-behavior and exhibit much lower write-amplification compared to LSM-Trees. 

\item MV-PBT has been implemented in PostgreSQL. The performance evaluation under HTAP workloads (CH-Benchmark \cite{chbench}) indicates that they improve the analytical throughput by 2x due to index-only visibility-checks, while improving the transactional throughput by 15\% compared to PostgreSQL's highly-optimized B\textsuperscript{+}-Tree. Under TPC-C MV-PBT performs 15\% better.  

\item MV-PBT has also been implemented in WiredTiger (MongoDB). The performance evaluation indicates approx. 40\% higher throughput under YSCB compared to WiredTigers highly-optimized LSM-Trees.

\end{itemize}

The rest of the paper is organized as follows. We motivate the missing \emph{version-awareness} and the need for \emph{index-only visibility-checks} in Section \ref{sec:motivation}, while Section \ref{sec:mvcc} provides some background on various multi-versioning aspects. The design and implementation of MV-PBT is described in detail in Section \ref{sec:mvpbt}, while the experimental evaluation is presented in Section \ref{sec:experimental_evaluation}. We. discuss related approaches in Section \ref{sec:related_work} and conclude in Section \ref{sec:conclusion}.

\section{Motivation}
\label{sec:motivation}

\begin{figure}[h]
	\centering
		\hspace{-20pt}
		\includegraphics[scale=1.35]{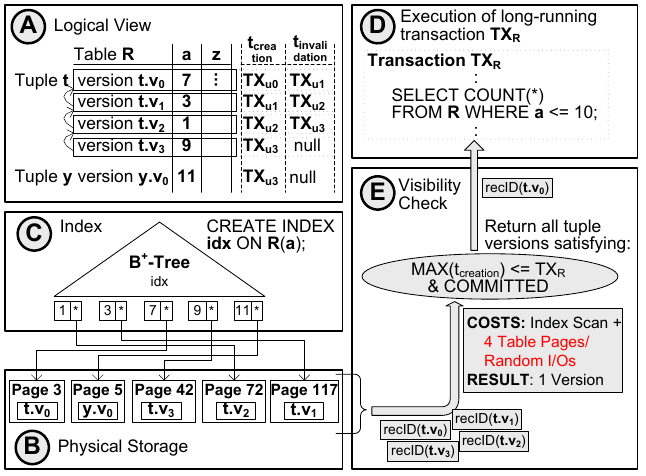}
		\vspace{-10pt}
	\caption{Index-Only Visibility-Check in Multi-Version DBMS: (a) logical tuples (t and y) of a table R and their versions; (b) the physical storage of these versions into database pages; (c) an index created over the table R must index all versions; (d) an index-scan retrieves all versions meeting the predicate, out of which  (e) the visibility-check returns the ones visible to calling transaction $TX_{R}$.}
	\label{fig:indexonly}
	\vspace{-5pt}
\end{figure}

In this section we give a more detailed perspective on the above issues of:
1) \emph{Version-obliviousness in indices}; 
2) \emph{missing index-only visibility-check}; and
3) \emph{I/O overhead}. 
Consider the example in Figure \ref{fig:indexonly}, which is a more detailed version of Figure \ref{fig:headpcit} with a conventional B\textsuperscript{+}-Tree. An initial transaction $TX_{U0}$ (not depicted) inserts tuple \emph{t} prior to $TX_{R}$, creating its initial version $t.v_{0}$. While $TX_{R}$ is running, multiple concurrent transactions $TX_{U1}$ \dots $TX_{U3}$ update tuple \emph{t} and each of them produces new versions of it ($t.v_{1}$\dots $t.v_{3}$). Only $TX_{U3}$ inserts tuple \emph{y} in its initial version $y.v_{0}$ in addition to creating $t.v_{3}$. Each tuple-version is a separate physical version record (Figure \ref{fig:indexonly}.A). It contains \emph{version-information}: the recordID of the predecessor version and two timestamps, \emph{$t_{creation}$} - the timestamp of the transaction that created that tuple-version; and \emph{$t_{invladiation}$} the timestamp of the transaction that invalidated it by creating a successor version. The invalidation-timestamp is \emph{null} if there is no successor. If a tuple gets deleted a special \emph{tombstone} version-record is inserted to mark the logical end of the chain. \emph{The version-information is only available on the version-record.}

Since version-records are independent physical entities they can be stored on any DB-page with enough free space. Figure \ref{fig:indexonly}.B depicts an example of the physical version-storage. \emph{An index on a table must contain index-entries for each committed version of every tuple for consistency.} Therefore, a B\textsuperscript{+}-Tree index \emph{idx} on attribute \emph{a} of table \emph{R} (Figure \ref{fig:indexonly}.C) should reflect all versions of each tuple of \emph{R}. \emph{Since the index is version-oblivious it contains no version-information, and treats each tuple-version as if it were a separate tuple.} 
Consequently, if $TX_{R}$ uses the index to count all tuples satisfying \emph{``a $\leq$10''} (Figure \ref{fig:indexonly}.D), the index scan will return the matching index entries (referencing versions $t.v_{0}$ \dots $t.v_{3}$). Now, each one of them must be checked for visibility, i.e. is it latest committed tuple-version prior to the start of $TX_{R}$. However, the necessary timestamps are available only on the version-records. Therefore, all of them are retrieved, at the cost of random I/Os.

Consider the example in Figure \ref{fig:indexonly}.D, C and E, the index-scan for the condition \emph{``a $\leq$10''} will return versions $t.v_{3}$, $t.v_{0}$, $t.v_{1}$ and $t.v_{2}$. Subsequently, they are read to extract the \emph{version-information} ($t_{creation}$ and $t_{invalidation}$ -- Figure \ref{fig:indexonly}.A) yielding four random I/Os. The visibility-check then determines the latest version committed prior to the start of $TX_{R}$, returning the recordID of $t.v_{0}$ and ignoring the rest. \emph{Since the index is version-oblivious and thus does not support index-only visibility-checks, the I/O costs amount to: COST(Index-Scan) + 1 random I/O for each matching tuple-version.} \emph{Especially for HTAP workloads this yields significant performance degradation depending on the length of the version-chains.}

\begin{figure}[h]
	\centering
	\includegraphics[width=\columnwidth]{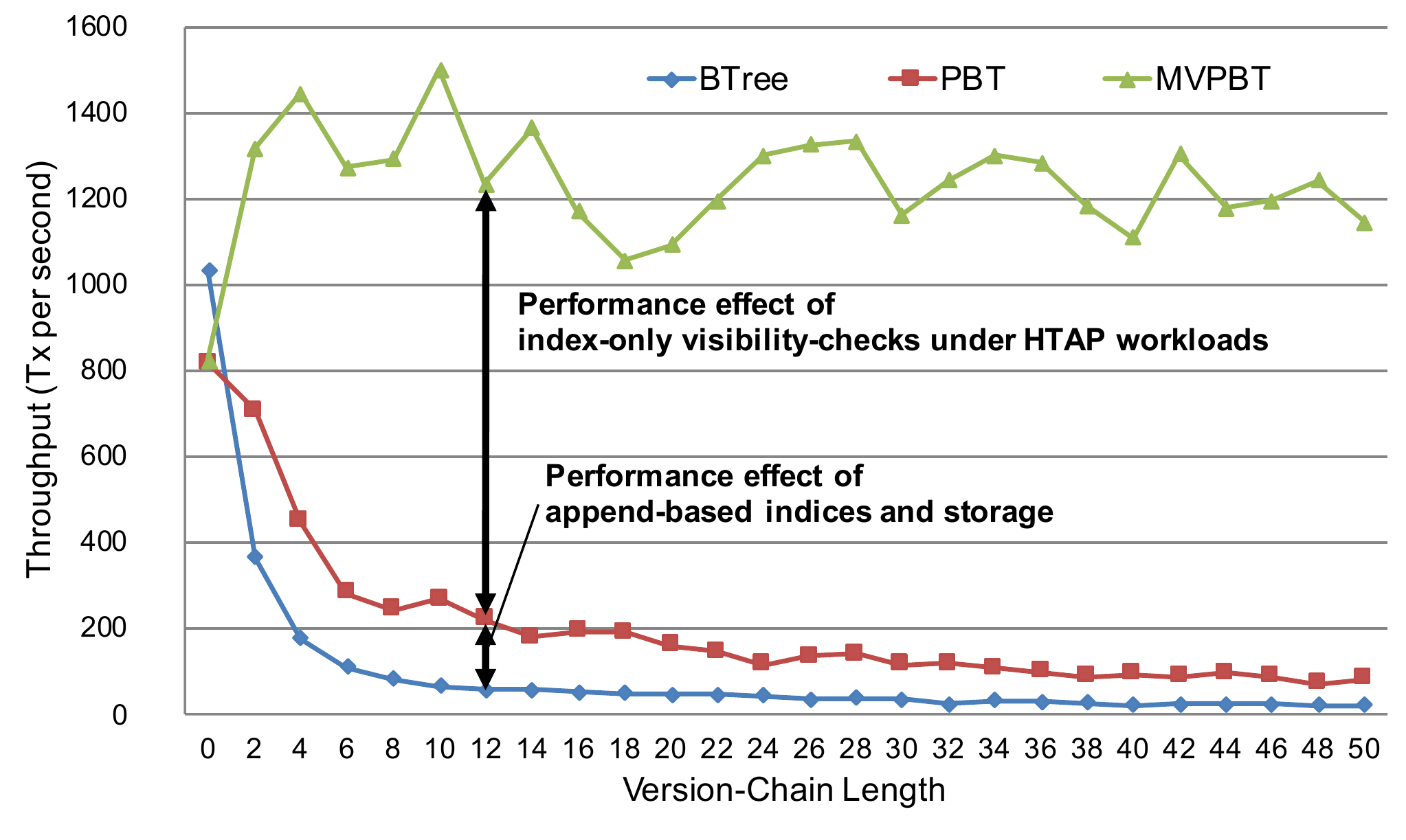}
	\vspace{-15pt}
	\caption{Performance Impact of Version Visibility Check.}
	\label{fig:chainlength}
	\vspace{-10pt}
\end{figure}

To quantify the combined effect, we designed a simple experiment with \emph{YCSB} and PostgreSQL. We run YCSB workloads A (update) and E (scan) combined,
performing frequent scans and updates. In parallel, we perform a point-query on a tuple every 30 seconds (simulating an HTAP workload). Additionally, we continuously increase the version-chain, by updating the tuple, until 50 versions are reached. In realistic HTAP settings, the amount of active versions can be as high as \emph{several hundred millions}, while analyses can take as long as 1000s \cite{Lee:HybridMVCC:GC:SIGMOD:2016}. 
The experimental results are shown in Figure \ref{fig:chainlength}. The highly-optimized B\textsuperscript{+}-Tree implementation in PostgreSQL performs better than \emph{MV-PBT} on a single tuple-version. However, as the version-chain length increases (6-8 versions) the performance drops rapidly to \emph{approx. 50 transactions/sec}, due to version-obliviousness and random I/O. Basic Partitioned B-Trees (PBT), are likewise \emph{version-oblivious}, but exhibit append-based write behaviour, avoiding in-place updates and perform therefore slightly better (\emph{approx. 150 tx/sec}). Due to its \emph{version-awareness} and support for \emph{index-only visibility-check} MV-PBT exhibits much higher and robust performance (\emph{approx. 1200 tx/sec}) with growing chain lengths.

\section{Background}
\label{sec:mvcc}
Multi-Version Concurrency Control (MVCC) is one of the most popular transaction management schemes and is used in most modern DBMS: Oracle, Microsoft SQL Server, HyPer, SAP HANA, MongoDB WiredTiger, NuoDB, PostgreSQL or MySQL-InnoDB, just to name a few. These DBMS  make different design decisions regarding the aspects described below.

\subsection{Version Storage}
\label{ssec:version_storage}
\begin{figure}[h]
	\begin{center}
		\includegraphics[width=\columnwidth]{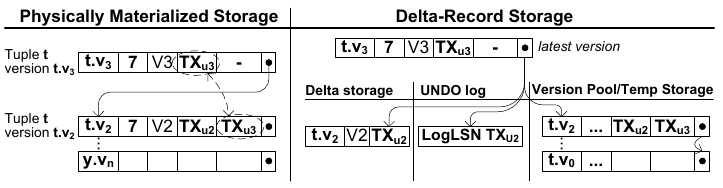}
		\vspace{-15pt}
 		\caption{Version Storage Alternatives}
		\vspace{-7pt}
	\label{fig:versionstorage}
	\end{center}	
\end{figure}

Under MVCC a logical tuple corresponds to one or more tuple-versions (Figure \ref{fig:indexonly}.A). They form a singly linked list, which represents a version chain. There are two possible physical representations of a tuple-version (Figure \ref{fig:versionstorage}): 
\emph{physically materialized} 
or 
\emph{delta-record based}.  
The former implies that each tuple-version record is stored physically materialized in its entirety and is in the focus of this paper. The latter implies that each modification of a logical tuple results in a delta-record, indicating the difference to another version (\`a la BW-Tree \cite{Levandoski:BWTree:ICDE:2013,Wang:BWTree:SIGMOD:2018}). The delta-records are connected and retrieved on demand by the DBMS storage manager to restore a tuple-version. \emph{Delta-record based} system designs typically store a single version (oldest or newest) in the main store and use a separate store for the delta-records, which may be the undo log (\`a la InnoDB) or a temporary version store (\`a la MS SQL Server). Both organizations can perform modifications in-place or out-of-place. Out-of-place updates with \emph{physically materialised} version-maintenance insert a new version-record in the base table. Based on the version ordering, additional modifications may be necessary to maintain logical timestamps or references.

Considering the characteristics of modern storage technologies, physically materialized version storage and out-of-place updates are preferable, due to lower write-amplification and the higher parallelism. Delta records tend to consume less space than materialized tuple-versions, but require additional processing and all predecessors or successors for tuple reconstruction.

\subsection{Version Ordering}
\label{ssec:version_order}
\begin{figure}[h]
	\vspace{-7pt}
	\begin{center}
		\includegraphics[width=.8\columnwidth]{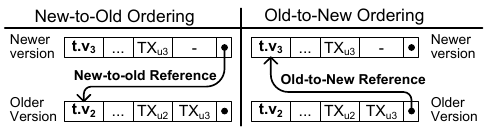}
		\vspace{-7pt}
		\caption{Version Ordering Alternatives}
		\vspace{-20pt}
	\label{fig:physical:version:ordering}
	\end{center}	
\end{figure}

The set of tuple-versions of a database tuple is organized as a singly linked list. There are two different ordering methods (Figure \ref{fig:physical:version:ordering}): \emph{old-to-new} and \emph{new-to-old}. 

\emph{Old-to-New ordering}: The entry-point is the oldest tuple-version in version chain and each version contains a reference (recordID) to its successor. A visibility-check must therefore process all successors, beginning from the oldest tuple-version. This behavior is beneficial for lookups of long-running analytical (OLAP) queries under HTAP workloads, where older tuple-versions are likely to be the visible ones. Alternatively, OLTP workloads mostly require the newest version and would need to process the whole version chain. \emph{New-to-Old ordering} implies that the entry-point is the newest tuple-version, which refers to its predecessor. Queries in the typically short OLTP transactions find the visible version very fast, but long-running OLAP queries may need to process several successors in version chain (Figure \ref{fig:chainlength}). \emph{In-place} and \emph{out-of-place} update strategies are are possible for both methods.

Considering the characteristics of modern storage technologies new-to-old ordering for physical version storage results in lower write-amplification and matches append-only storage. All other approaches require in-place updates.

\subsection{Version Invalidation Model}
\label{ssec:invalmodel}
\begin{figure}[h]
	\begin{center}
		\includegraphics[width=.9\columnwidth]{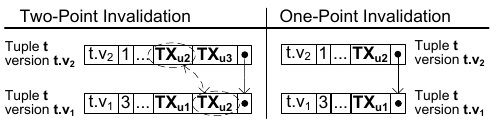}
		\vspace{-7pt}
		\caption{Version Invalidation Model}
		\vspace{-15pt}		
	\label{fig:invalidation:model}
	\end{center}	
\end{figure}

Under MVCC a version is said to be invalidated whenever a successor version exists. There are two possible invalidation models \cite{Gottstein:Diss:2016} (Figure \ref{fig:invalidation:model}). 
First, \emph{two-point invalidation} is the state-of-the-art model, where the creation timestamp of the successor version is also placed as invalidation timestamp on the predecessor.  \emph{Two-point invalidation} works well with old-to-new ordering. However, with new-to-old ordering, the invalidation timestamp must be set on the predecessor version, yielding an in-place update and possibly a random write. 
Second, with \emph{one-point invalidation} \cite{gottstein:siaschains}, the existence of a successor implicitly invalidates the predecessor and all version-records contain only the creation timestamp. \emph{One-point invalidation} matches well new-to-old ordering, the use of indirection layer (VIDs, and entry-points) as well as append-based storage.

\subsection{Garbage Collection}
\label{ssec:gc}
Under MVCC modifications of a tuple result in the creation of a new tuple-version. Old tuple-versions become obsolete, if they are no longer visible to any of the active transactions. Therefore, some form of \emph{version GC} is necessary to reclaim space and can improve performance. However, GC causes performance spikes (as it interferes with foreground I/O), reduces concurrency (as some form of locking is required) and increases write-amplification on secondary storage. 
GC \cite{pavlo:mvcc_eval} can be performed on transaction \cite{Lee:HybridMVCC:GC:SIGMOD:2016}, tuple and index levels \cite{Levandoski:BWTree:ICDE:2013,Wang:BWTree:SIGMOD:2018}. 
Index-level GC (Section \ref{ssec:mvpbt_in_memory_gc}) purges index entries, resulting from index updates, maintenance or tuple-level GC.

\subsection{Version/Index-Record Referencing}
\label{ssec:index_management}
\begin{figure}[h]
	\begin{center}
		\includegraphics[width=.85\columnwidth]{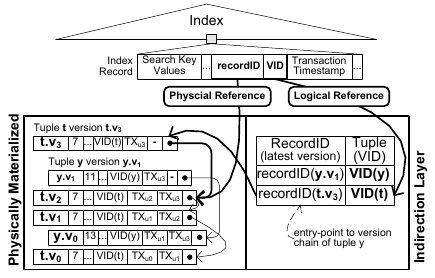}
		\vspace{-7pt}		
		\caption{Version/Index-Record Referencing}
		\vspace{-7pt}		
	\label{fig:refernces}
	\end{center}	
\end{figure}

There are two possibilities to map index records to tuple-versions in base tables (Figure \ref{fig:refernces}). \emph{First}, classical physical references (recordIDs) can be used. Thus, the latest tuple-version in base tables (entry-point in the version chain) can be accessed directly, but changes to the latest version or its location result in index-record modifications. Such changes comprise: creation of a successor-version; storage management and physical movement (as in append storage) or garbage collection. \emph{Second}, an indirection layer with logical references can be employed. Each tuple-version is augmented with an unique tuple-identifier (Virtual Tuple Identifier -- VID), which is also stored in the index records. An index operation resolves the VID using a mapping table (indirection layer) to locate the physical entry-point. \emph{An indirection layer can reduce index maintenance costs for in-place and out-of-place updates, 
but requires additional structures and processing.} 

Traditional index designs use physical references and contain no version-information, which tends to increase index maintenance overhead as well as the visibility check costs for lookups and scans. Alternatively, modern index-structures (BW-Tree) use an indirection layer, but contain no version-information and support no index-only visibility check. 
This can cause massive read amplification for mixed workloads. An optimal index structure should reduce write amplification and return only references to tuple-versions that are visible to a transaction snapshot. A MV-PBT uses physical or logical references, is version-aware and produces append-only sequential write pattern.

\subsection{Discussion} 
\label{ssec:discuss}
We have outlined some relevant design decisions for storing tuple-versions in multi-version DBMS. Modifications are preferably stored as \emph{physically materialised tuple-versions} in base tables, rather than deltas, due to tuple reconstruction costs. Moreover, this enables direct access to each tuple-version from additional access paths. \emph{Out-of-place updates} reduce write amplification to secondary storage. 
Garbage collection is required for space reclamation, but brings additional complexity to data structures.  

A \emph{new-to-old} version ordering requires index maintenance for every new tuple-version, because the entry-point of the version chain for that tuple changes. A logical indirection layer ensures fast lookups by efficiently returning entry-point to a version chain and reduces index maintenance effort. \emph{New-to-old} ordering is beneficial for OLTP and speeds up visibility-check as the latest version ist typically the visible one, yet older versions may require slow reconstruction. Alternatively, \emph{old-to-new} ordering is supports long-running OLAP operations and visibility-check in HTAP settings, as the oldest version is directly accessible. Yet, modifications and maintenance may suffer low performance.

Indices for mixed workloads and large datasets should rather return visible tuple-versions. Alternatively, traditional index structures only return version candidates, which have to be subsequently verified, fetching version-records from base tables by performing random I/O. 

\section{Characteristics of modern Hardware Technologies}
\label{sec:modern_hardware}
\begin{figure}[h]
	\centering
		\hspace{-20pt}
		\includegraphics[scale=.7]{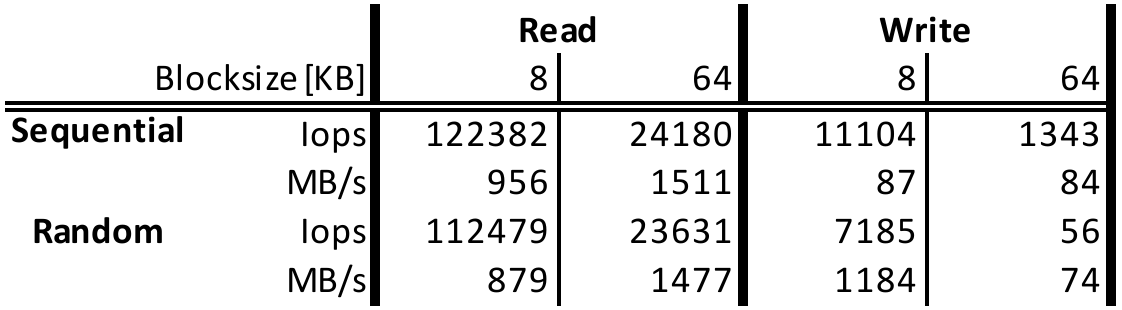}
		\vspace{-10pt}
	\caption{I/O Characteristics of Intel DC P3600 SSD.}
	\label{fig:p3600}
	\vspace{-10pt}
\end{figure}

Modern database storage management needs to address the characteristics of semiconductor storage technologies \cite{7113423}. Consider Figure \ref{fig:p3600} depicting the I/O characteristics of the enterprise Flash storage used in the evaluation. Typical index search operations result in large amount of small (8K) random reads. Hence, optimize for read IOPS and sequential writes ($\geq$64K).
We derive the following tradeoffs for the I/O behaviour of MV-PBT: 
(a) transform random writes in sequential writes with higher granularity (MB); and 
(b) trade sequential writes for complex and possibly random reads with higher parallelism and smaller granularity (KB). Thus, append-based storage managers are beneficial for the base tables \cite{gottstein:siaschains,Gottstein:Diss:2016}. Write-sequentialization is therefore necessary for indices, and MV-PBT support it intrinsically, like LSM-Trees. 

\section{Multi-Version Partitioned B-Trees}
\label{sec:mvpbt}

\begin{figure}[h]
	\begin{center}
		\includegraphics[width=\columnwidth]{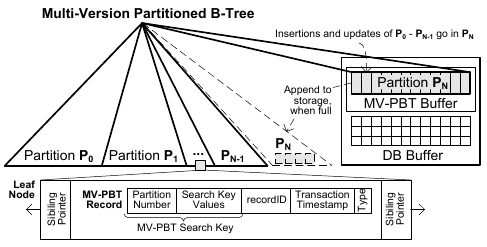}
		\caption{Structure of a Multi-Version Partitioned B-Tree.}
		\vspace{-5pt}
		\label{fig:mvpbt}
		\vspace{-5pt}
	\end{center}
\end{figure}

Multi-Version Partitioned B-Trees (MV-PBT) -- Figure \ref{fig:mvpbt} -- are based on Partitioned B-Trees (PBT), introduced by Goetz Graefe \cite{graefe:sortingindexingpbt,graefe:pbt_guide}. PBT in turn represent an enhancement on traditional B\textsuperscript{+}-Trees\cite{Bayer:BTREE:SIGMOD:1970}. PBT (and MV-PBT) create index partitions based on an artificial, leading key-column -- \emph{the partition number}. All index-entires in a partition have the same partition number in the search key. PBT (and MV-PBT)  utilize a portion of the database buffer (\emph{partition buffer}) to host the latest partition $P_{N}$, where insertions and updates to existing partitions ($P_{0}$ \dots $P_{N-1}$) are placed. Updates to existing index entries are treated as \emph{replacement records} to avoid in-place updates. Once $P_{N}$ gets full, a MV-PBT partition is appended to persistent storage and becomes immutable.


Regular MV-PBT records comprize of a \emph{partition number}, its \emph{search key columns}, and a \emph{recordID (set)}. Furthermore, MV-PBT index records contain version-information: \textit{logical transaction timestamp} for validation or invalidation of the tuple-version and optionally an \emph{unique virtual identifier} (indirection layer). 
Each partition number value identifies a single partition. Partition numbers are unique, monotonically increasing,  two-byte integer values. This enables the MV-PBT to maintain partitions within one single tree structure in alphanumeric sort order. The partition number is an artificial column and is therefore transparent to higher database layers. Each MV-PBT maintains partitions independent of other MV-PBTs. Partitions appear and vanish as simple as inserting or deleting records. They can be reorganized and optimized on-line in system-transaction merge steps, depending on the workload. Partitions can support additional functionalities, like bulk loads or can serve as multi-version store\cite{graefe:sortingindexingpbt}.

MV-PBTs write any modification of index records exactly once -- upon eviction of a partition, except for later reorganization or garbage collection operations. This is realized by forcing sequential writes of all leaf nodes in a partition (Figure \ref{fig:mvpbt}). Leaf nodes of modifiable main memory partitions are stored in a separate buffer cache -- the \emph{MV-PBT Buffer}. This area is shared for all MV-PBT indices in the database. Once the MV-PBT Buffer gets full, a victim MV-PBT is selected  and its $P_{N}$ is written to secondary storage. The MV-PBT Buffer is managed by a special replacement policy, giving active partitions the chance to grow (Section \ref{ssec:mvpbt_buff_mgmnt}). 

\subsection{MV-PBT Record Types}
\label{ssec:mvpbt_rec_type}
Persistent index partitions are immutable. Direct modification-operations are forbidden. Therefore, modifications to existing index-records as well as insertions are placed in the buffered partition $P_{N}$. To handle this behavior MV-PBT introduces new index-record types. Currently the following are defined.

\begin{figure}[h]
	\begin{center}
		\includegraphics[width=\columnwidth]{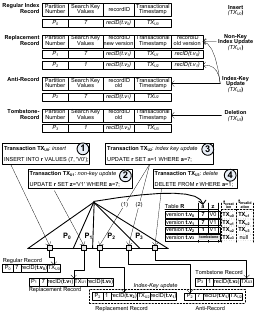}
		\vspace{-10pt}
		\caption{MV-PBT Index-Record Types and Their Use: MV-PBT record format (top), an example including a sequence of transactions and their index records (bottom).}
		\label{fig:recordtypes}
		\vspace{-20pt}
	\end{center}
\end{figure}

{\bf Regular Index Records} 
are created upon the insertion of new tuples.  The \emph{partition number} of the newest MV-PBT partition $P_{N}$ is inserted together with the search key values. The \emph{recordID} (pageID and slot) of the newly inserted tuple-version is included as well as the \emph{transaction timestamp} of the inserting transaction (Figure \ref{fig:recordtypes}). The latter is essential for index-only visibility-checks. 
For example, transaction $TX_{U0}$ (Figure \ref{fig:recordtypes}) inserts a new tuple (\emph{t}), in its initial version (\emph{$t.v_{0}$}), causing the creation of a \emph{regular index record} in partition $P_{0}$.

{\bf Replacement-Records}
result from tuple-updates on \emph{non-index key columns} on existing index-entries. Such updates yield a new tuple-version that becomes the new chain entry-point, which needs to be reflected in the index. Although the index-record for the previous version has not changed (non-index-key update) the version-information and recordID of the new version need to be replaced. However, this is not possible, if the index-record is already in an immutable partition ($P_{0}$\dots $P_{N-1}$). Therefore a \textit{replacement record} is inserted in the newest partition $P_{N}$ to logically replace the old one with the recordID and the version-information. 
The \textit{Replacement Record} (Figure \ref{fig:recordtypes}) contains: the recordID of the new version, its creation-timestamp as well as the recordID of the predecessor version. Hence the record includes some "anti-matter" \cite{graefe:sortingindexingpbt} (recordID) invalidating the predecessors as well as some "matter", i.e. recordID and timestamp if the new version. 
For example, transaction $TX_{U1}$ (Figure \ref{fig:recordtypes}) updates the attribute \emph{z} of the previously inserted tuple (\emph{t}), producing a new version (\emph{$t.v_{1}$}). Although the index-key \emph{7} remains unchanged, the version-information of (\emph{$t.v_{1}$}) has to be updated, causing the creation of a \emph{replacement-record} in partition $P_{1}$.

{\bf Anti-Records} 
are required for updates on index-key attributes and are always used in combination with \emph{replacement records} in the same partition. 
If the index-key of an existing index-record (in the immutable partitions) gets updated, 
MV-PBT inserts a combination of an \emph{anti-record} and a \emph{replacement record}. \emph{Anti-records} are pure "anti-matter" as they mark extinction of the old index record (from partitions $P_{0}$\dots $P_{N-1}$), whereas the simultaneously inserted \emph{replacement record} represents the new "matter" and reflects the new index-key and the new version-information. The anti-record and the replacement records are inserted in $P_{N}$ and are placed according to the sort-order of the search-key value.
An anti-record contains the recordID of the old version, together with its search key and the transaction timestamp of the updating transaction (Figure \ref{fig:recordtypes}). 
For example, transaction $TX_{U2}$ (Figure \ref{fig:recordtypes}) updates the indexed attribute \emph{a} of the previously inserted and updated tuple (\emph{t}), producing new version (\emph{$t.v_{2}$}), modifying the index-key from \emph{7}  to \emph{1}. 
The \emph{anti-record} (marking the extinction of the replacement-record from partition $P_{1}$) reflects the recordID of the predecessor version (\emph{$t.v_{1}$}), contains its index-key values (7) and the transaction-timestamp of the current updating-transaction ($TX_{U2}$). The simultaneously inserted replacement record reflects the new and updated value of the search key (i.e. \emph{1}), the recordIDs of the old and the new tuple-versions (\emph{$t.v_{1}$ and $t.v_{2}$}) as well as the transaction-timestamp of $TX_{U2}$. Since the index records are kept in sort order of the search-key values within a partition (as in a B-Tree) the replacement record is placed first in order, followed by the anti-record.

{\bf Tombstone-records} 
indicate the deletion of a tuple. If a tuple is logically deleted, it does not become erased immediately in MV-DBMS, because it could be visible to a concurrent transaction. Rather a tombstone tuple-version record is inserted in the DB, which needs to be reflected in the MV-PBT index.
\textit{Tombstone-records} are similar to \textit{Anti Records} in that they represent pure "anti-matter", marking the extinction of the whole tuple-version chain. The difference is that if a tombstone-record is visible to a transaction, no further tuple-version belonging to this chain can be visible, even no \textit{replacement record}.
\textit{Tombstone-records} (Figure \ref{fig:recordtypes}) contain the recordID of the latest tuple-version and the transaction-timestamp of the deleting transaction.

For example, transaction $TX_{U3}$ (Figure \ref{fig:recordtypes}) deletes tuple (\emph{t}), creating a tombstone-version (\emph{$t.v_{3}$}) in the DB. Therefore a tombstone record is inserted in partition $P_{3}$ with the recordID of the deleted tuple-version  $t.v_2$, reflecting deletion of the whole version chain $t.v_{2} \rightarrow t.v_{1} \rightarrow t.v_{0}$.



\subsection{MV-PBT Operations}
\label{ssec:mvpbt_ops}

In the following we describe the index operations in a MV-PBT:
\begin{itemize}[leftmargin=*,noitemsep,nolistsep]

\item \emph{Insert Operations} are only performed in $P_{N}$. An \emph{insertion} yields the creation of an \emph{regular index-record} in $P_{N}$ with the recordID of the newly created tuple-version and the timestamp of the creating transaction. The insertion traverses the buffered partition $P_{N}$ and places the new index record according to the alphanumeric sort-order of the search-key (ordering issues are described in Section \ref{ssec:mvpbt_version_order}). The MV-PBT buffer management strategy (Section \ref{ssec:mvpbt_buff_mgmnt}) guarantees sufficient space for the insertion and possible maintenance. In case of an \emph{non-unique index} the insertion is performed directly. Alternatively, given a \emph{unique index} a lookup operation (see Search and Scan) is performed first to guarantee the non-existence of the new index-key. 

\item \emph{Update Operations} are performed in different ways. If a transaction modifies a tuple-version in a way that a non-index-key attribute is changed (\emph{non-key update}) a new tuple-version is created and its version-information needs to be reflected in the index. In case of \emph{non-key updates} MV-PBT inserts a \emph{replacement-record} in $P_{N}$ (Figure \ref{fig:recordtypes}), containing the version-information (recordID and timestamp) of the modifying transaction. By doing so, it logically \emph{replaces} the index-record, which is located in an older partition, and reflects the predecessor version. 

Alternatively if the modifying transaction updates an index-key attribute (\emph{index-key update}) a \emph{replacement record} as well as an \emph{anti-record} are inserted in $P_{N}$ (ordering issues are described in Section \ref{ssec:mvpbt_version_order}). The former reflects the new and modified index-key value in the new tuple-version, the latter indicates the extinction of the old index-record, reflecting the index-key value of the predecessor version. In case of an \emph{unique index}, the MV-PBT first performs a lookup to ensure the non-existence of the new key-value. 

\item \emph{Delete Operations} cause the insertion of a  \emph{tombstone record} in $P_{N}$. If a transaction deletes a logical-tuple a tombstone version is created indicating the deletion of the whole version-chain, to transactions to which it is visible. Analogously MV-PBT inserts a \emph{tombstone record} to indicate the extinction of all index-records corresponding to the version chain. Ordering issues are described in Section \ref{ssec:mvpbt_version_order}.


\item \emph{Search and Scan Operations} process partitions in reverse order from $P_{N}$ to $P_{0}$. Filter techniques such as \emph{Partition Range Keys}, \emph{Minimum Transaction Timestamp} or \emph{Bloom- and Range Filters} (Section \ref{ssec:mvpbt_filters}) are needed for selecting the predeceasing partition which may contain an index record, matching the search conditions (Algorithm \ref{alg:getnext}). The search conditions are extended to match the format of a MV-PBT -- the partition number is prepended to the first search key column. A regular root-to-leaf traversal operation is performed and the cursor is positioned. Afterwards, the next matching index record is requested and checked for visibility (Section \ref{ssec:mvpbt_visibility_check}). This process is repeated until an index record visible to current transaction is found, and can be returned together with the respective \textit{recordID}. \textit{Partition number} and \textit{timestamp} are transparent for higher database layers and become removed. Index records of most recent tuple versions are found and processed first, due to index-record ordering (Section \ref{ssec:mvpbt_version_order}), which is very beneficial for simple search conditions, like point lookups. 

\emph{Complex scan} operations (Algorithm \ref{alg:scan}) build a set of all matching index records,  spreading all MV-PBT partitions. Every partition is pre-selected by filter techniques and processed from $P_{N}$ to $P_{0}$. Traversal operations benefit from commonly buffered higher levels of the tree-structure. Matching index records of any record type in a partition are processed by the index-only visibility-check. Visible index records are added to the result set without  \textit{partition number} and \textit{timestamp} in regular sort order. If no further index record matches the scan conditions, the algorithm proceeds with the predeceasing partition. 
Finally, the result set is returned. It is filled with all index records (including \textit{recordIDs}), matching the scan and visibility conditions of the calling transaction. 

A single scan process without rechecking for concurrent modifications in $P_N$ is sufficient, due to transaction snapshots -- concurrent modifications in $P_N$ are invisible, anyways. Expensive retrieval of version-records in base-table (\textit{random read I/O}) for \textit{version-information} is avoided. In case of selection of non-index attributes, the recordID indicates the location of version-record in base-tables, which can be directly accessed.


\begin{algorithm}
\caption{MV-PBT Search}
\label{alg:getnext}
\begin{algorithmic}[1]

 \Function{search}{Search conditions ${|attr_{val,cond}|, ...}$} \\ {{\bf Output:} $IndexRecord$}
	\While{\Call{hasNext}{\ }}
	\State Let $idx\_record \gets \Call{next}{\ }$ \Comment fetch next index record
	\If{\Call{VisCheck}{$idx\_record$} \textbf{equals} $VISIBLE$}
	\State \Return \Call{set\_return\_format}{$idx\_record$} \indent \indent \indent \Comment{hide $partition number$ and $timestamp$}
	
	\EndIf
	\EndWhile
	\While{$part \gets \Call{previousPartition}{part}$}
	\If{${|attr_{val,cond}|} \in part.filter$}
	\State Let $|skeys_{part}| \gets \ \ \ \ \ \ \ \ \ \ \ $ 
	\indent  \indent \indent \indent \Call{form\_rec}{$part,{|attr_{val,cond}|}$}
	\State \Call{traverse}{$|skeys_{part}|$} 
	\State \Return \Call{search}{\ }
	\EndIf
	\EndWhile
	\State \Return $\emptyset$
 \EndFunction
\end{algorithmic}
\end{algorithm}

\begin{algorithm}
\caption{MV-PBT Scan}
\label{alg:scan}
\begin{algorithmic}[1]
	
	\Function{scan}{Scan conditions ${|attr_{val,cond}|, ...}$} \\ {{\bf Output:} ResultSet of $|IndexRecords|$}
	\State $part \gets \emptyset$ \Comment{\Call{previousPartition}{} returns $P_N$ for $\emptyset$}
	\While{$part \gets \Call{previousPartition}{part}$}
	\If{${|attr_{val,cond}|} \in part.filter$}
	\State Let $|skeys_{part}| \gets \ \ \ \ \ \ \ \ \ \ \ $ 
	\indent  \indent \indent \indent \Call{form\_rec}{$part,{|attr_{val,cond}|}$}
	\State \Call{traverse}{$|skeys_{part}|$} 
	\EndIf
	\While{\Call{hasNext}{\ }}
	\State Let $idx\_record \gets \Call{next}{\ }$ \Comment neighbor in BTree
	\If{\Call{VisCheck}{$idx\_record$} \textbf{equals} $VISIBLE$}
	\State |$IndexRecords$|.\Call{add}{}( \indent \indent \indent \indent\Call{set\_return\_format}{$idx\_record$}) 
	
	\EndIf
	\EndWhile
	\EndWhile
	\State \Return $|IndexRecords|$
	\EndFunction
\end{algorithmic}
\end{algorithm}

\end{itemize}

\subsection{MV-PBT Index-Record(Version) Ordering}
\label{ssec:mvpbt_version_order}

The version/partition-placement in MV-PBT given by modification, search and scan algorithms of MV-PBTs. \emph{Index-records of predecessor versions are likely to be located in lower-numbered partitions, successors in higher-numbered ones (Figure \ref{fig:recordtypes}).} This however necessitates multiple memory partitions for a MV-PBT.


To address such issues the current MV-PBT design uses a single main-memory partition $P_{N}$ for each MV-PBT. \emph{However, for index-records with the same index-key it is mandatory that records for newer/successor versions are always placed before index-records for older/predecessor versions in $P_{N}$.}
In other words the \emph{primary sort-order} of the index-records in a $P_{N}$ is on the search-key (mostly descending), however all records with the same search-key are sorted in inverse \emph{secondary sort-order} (mostly ascending) on the transactional timestamp. 

Search and scan operations traverse partitions \emph{backwards}: starting from buffered partition $P_{N}$ (i.e. $P_{N} \rightarrow P_{N-1} \dots \rightarrow P_{0}$).
Yer, given the above ordering, index-records of newer tuple-versions, matching the search predicates, are processed \emph{first} in forward direction (i.e. in descending timestamp-order). Only then the next lower-numbered partition is traversed and processed. \emph{This is how MV-PBT ensures that in a search and scan operation, newer versions can always be found before older ones in the same partition, and across partitions. }

\begin{figure}[h]
	\begin{center}
		\includegraphics[width=\columnwidth]{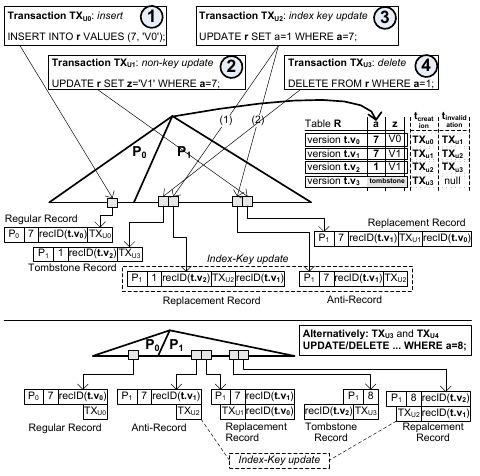}
		\vspace{-15pt}
		\caption{MV-PBT Index-Record Ordering.}
		\label{fig:versionordering}
		\vspace{-15pt}
	\end{center}
\end{figure}

Consider for example Figure \ref{fig:versionordering}, where we have only two partitions and index-records reflecting updates to the same tuple go to $P_{1}$, and contrast to Figure \ref{fig:recordtypes}, where all index-records with higher-timestamps are placed in higher-numbered partitions. Observe that the index-records in $P_{1}$ (Figure \ref{fig:versionordering}) appear in their primary-order (on the search key), i.e. records with search-key 1 precede those with 7. Observe also that the \emph{tombstone record} with key 1 precedes the \emph{regular record} as a result of the \emph{secondary sort-order} since $timestamp(TX_{U3}) > timestamp(TX_{U2})$. 

\subsection{MV-PBT Index-Only Visibility-Check}
\label{ssec:mvpbt_visibility_check}
MV-PBT is \emph{version-aware} and supports \emph{index-only visibility-check}, i.e. it returns a set of index records matching the search condition and \emph{visible} to the calling transaction. In doing so, MV-PBT avoids the expensive retrieval of base-table version-records to extract their \emph{version-information}. 

The \emph{index-only visibility-check} (Algorithm \ref{alg:idx_only_vchck}) is inherently supported by the data structure. MV-PBT index records (Section \ref{ssec:mvpbt_rec_type}) contain version-information and define modifications and recordIDs of tuple-versions. The respective \emph{index-record ordering} is essential to scans (Section \ref{ssec:mvpbt_version_order}), whereby records indicating the invalidation of a tuple-version are guaranteed to be placed before the \emph{``validating''-records} for a given transactional timestamp. 

Index records of any type, matching the search-conditions are processed by the visibility check. They are \emph{invisible} to a transaction, if: 
\begin{enumerate}[leftmargin=*,noitemsep,nolistsep]
\item [(a)] the index record is \emph{flagged for garbage collection}; 
\item [(b)] the transaction timestamp of the index-record is \emph{greater than} the timestamp of the calling transaction; or \\
the transaction corresponding to the index-record timestamp is \emph{concurrent} to the calling transaction;
\item [(c)] visible record with \emph{anti-matter} for the recordID (anti-matter, replacement- and tombstone-records) was already encountered (in this case also checked for GC); or 
\item [(d)] the index record is either a \emph{tombstone record} or an \emph{anti-record}. 
\end{enumerate}
An additional visibility-check by processing the version chain in base table is not required. Skewed updates on tuples do not lower the performance of the index-only visibility check, due to well performing garbage collection and well-cached version-chains in the main-memory partition $P_{N}$.
%
%
%
%
%
%
%

\begin{algorithm}[!h]
\caption{MV-PBT Index-Only Visibility-Check}
\label{alg:idx_only_vchck}
\begin{algorithmic}[1]
\Function{VisibilityCheck}{	$idx\_record$} \\
	\textit{{\bf input: }} $idx\_record$ at current scan position\\ 
	{\textit{{\bf output:}} $BOOL\_VISIBLE$}

	\State Let $anti\_map \gets Map$ of $(recID | TS)$ \Comment{anti-matter} 
	\If{\Call{IS\_SET}{$idx\_record, FLAG\_GC$}}
	\State \Return $INVISIBLE$
	\EndIf
		\If{\textbf{not} \Call{precedes}{$idx\_record.ts,CurrentTxId$} \textbf{OR} 
		\indent $\ \ \ \ \ \ \ $ \Call{isConcurrent}{$idx\_record.ts,CurrentTxId$}}
	\State \Return $INVISIBLE$
	\EndIf
		\If{$ts_{anti} \gets anti\_map.\Call{get}{idx\_record.recID_{matter}}$ \indent \textbf{and} $\Call{precedes}{idx\_record.ts,ts_{anti}}$ }
		\State \Call{checkForGC}{$idx\_record$}
	\State \Return $INVISIBLE$
	\EndIf
	\If{\Call{IS\_SET}{$idx\_record, FLAG\_ANTI\_MATTER$}}
	\State $anti\_map.\Call{put}{idx\_record.recID_{anti}, idx\_record.ts}$
	\EndIf
		\If{\Call{IS\_SET}{$idx\_record, FLAG\_MATTER$}}
	\State \Return $VISIBLE$
	\EndIf
	\State \Return $INVISIBLE$ 
\EndFunction
\end{algorithmic}
\end{algorithm}

\subsection{MV-PBT Buffer Management}
\label{ssec:mvpbt_buff_mgmnt}
MV-PBT accumulate modifications to persistent partitions in the latest partition $P_{N}$, which is held in the MV-PBT partition buffer (Figure \ref{fig:mvpbt}). All MV-PBT indices place their respective $P_{N}$ in the MV-PBT buffer which rises the question of the proper buffer management strategy. Well-known replacement policies (like LRU or ARC) are not suitable for managing the set of leaf nodes contained in the respective $P_{N}$ as well as different $P_{N}$. The MV-PBT buffer should (a) only evict partitions as a whole instead of individual pages (like in LRU) to achieve sequential write patterns; and (b) give partitions of update intensive indices a fair chance to grow, and balance it across all indices. Remember that MV-PBT read operations place persistent partition nodes in the main/shared DB-Buffer. 
MV-PBT buffer-management strategy can be summarized as follows. Whenever the buffer-size threshold is reached the MV-PBT buffer manager selects the largest  partition of all indices as a \emph{victim} for eviction. Smaller, less update-intensive partitions are frequently evicted to avoid imbalanced number of partitions per MV-PBT and shrinking partition sizes.

The \emph{eviction process} (Algorithm \ref{alg:mvpbt:eviction}) can be summarized as follows . A new partition numbered $P_{N+1}$ (initially $P_{N+2}$) is created for ongoing modifications. The current victim partition $P_{N}$ becomes \emph{immutable} and is scanned, as following operations are performed cooperatively and latch-free, piggybacking that \emph{scan}. 
\begin{enumerate}[leftmargin=*,noitemsep,nolistsep]
\item Version-chains are built from the Scan-ResultSet record using their timestamps and RecordIDs and creating a temporary VID for each chain. While doing that obsolete index-records (parts of the version-chain) are detected and marked for garbage collection.

\item \emph{Garbage Collection} is performed on the marked records (no longer needed/invisible records are removed) and the result is \emph{written out} to new leaf nodes. 

\item During this process index-records and leaf nodes are transformed to an on-disk format, whereby prefix-truncation, compression and encoding as well as dense-packing (Section \ref{ssec:mvpbt_filters}) are performed. Furthermore, the partition number of each index record is decremented from $P_{N}$ to $P_{N-1}$. Now $P_{N-1}$ is a separate partition, which is yet unknown in the MV-PBT partition metadata. The process resembles a leaf-build in PostgreSQL: having full pages the intermediary index nodes can be built on top. Concurrent, lookups and scans are still performed on the old $P_{N}$ nodes.

\item In parallel, well-sized \emph{(prefix-) bloom filters} are created (Section \ref{ssec:mvpbt_filters}). 

\item Dense-packing, compression and read-optimizations are performed to higher level intermediary nodes, resembling to a bottom-up build. All nodes are \emph{sequentially written out}. 

\item Finally, $P_{N-1}$ is added to the MV-PBT partition metadata. The old $P_{N}$ leaf nodes, on which concurrent non-blocking reads had been executing, are detached from the MV-PBT and are freed for reuse.
\end{enumerate}

\begin{algorithm}
\caption{MV-PBT Partition Eviction}
\label{alg:mvpbt:eviction}
\begin{algorithmic}[1]
\Function{evict} {$|P_{N}|$} \\
	\textit{{\bf Input: }} set of $P_{N}$ in MV-PBT buffer
	\State Let $p_{evict} \gets $\Call{SelectEvictionVictim}{$|P_N|$}
	\State Add $Partition\ p_{evict+1}$ to B$^+$-Tree $PartitionsList$
	\State \Call{SET}{$p_{evict}, FLAG\_IMMUTABLE$} 
	\State Let $recordSet \gets \Call{scanRecords}{p_{evict}}$
	\State \Call{garbageCollectionP3}{$recordSet$}
	\State $worker1.$\Call{loadAndFlush}{$p_{evict}.pNo-1, recordSet$}
	\State $worker2.$\Call{createFilters}{$p_{evict},recordSet$}
	\State \Call{wait}{\ }
	\State Let $p_{evict\_new} \gets $\Call{decrementPartitionNumber}{$p_{evict}$} 
	\State \Call{detatchAndFree}{$p_{evict}$}
\EndFunction
\end{algorithmic}
\end{algorithm}
\subsection{MV-PBT Partition Garbage Collection}
\label{ssec:mvpbt_in_memory_gc}
Mixed workloads with high update-rates result in massive amount of tuple-versions, which need to be garbage collected once a long-running reading/analytical query completes \cite{Lee:HybridMVCC:GC:SIGMOD:2016}. Same is true for the corresponding index-records. With high probability these records are located in the main-memory partition $P_{N}$ of a MV-PBT due to their temporal locality.
Therefore, we implemented a cooperative page-level garbage collection (GC) for $P_{N}$. 

\noindent \emph{Phase (1):} The GC \emph{piggybacks} regular index-scans to identify index-records of versions, that are not visible to any active transaction (\emph{cutoff-transaction}). As a page is already latched (shared), the following checks a performed on each record: (a) comparison with the lowest active transaction timestamp and if lower, mark predecessors as \emph{victim-versions} for GC; (b) if higher, but a successor exists, mark all predecessors as victims for GC. In both cases, a \emph{hasGarbage} flag is set in the page header (no exclusive latch required). This step also \emph{piggybacks} the in-memory structures of the scan and index-only visibility check algorithms. Records with \emph{anti-matter} (anti-matter, replacement and tombstone records) require special attention, as they are still required for invalidation of predecessors.  Hence the \emph{anti-matter} record with the highest timestamp smaller than \emph{cutoff} transaction timestamp must not be garbage collected. Index-record ordering (Section \ref{ssec:mvpbt_version_order}) supports GC  while scanning, since successors are mostly processed first. 

\noindent \emph{Phase (2):} Update operations check the \emph{hasGarbage} flag in page header. If set they first set the \emph{recordID} of the oldest required record with \emph{anti-matter} (anti-matter, replacement and tombstone records) to the recordID of the oldest victim-version of that chain on the page. Next, GC victims are removed on that page, the space is reclaimed and only then the update operation proceeds. This behavior saves memory, speeds up scans and visibility checks as well as reduces index maintenance operations (split).

\noindent \emph{Phase (3):} To handle version-chains spanning several pages, and for final cleanup before partition eviction the whole partition is scanned and the version chains (based on timestamps and records) are built in memory. This scan is also piggybacked for filter creation and dense-packing (Section \ref{ssec:mvpbt_filters}). Before switching to sibling page, obsolete versions are removed after updating invalidation reference and in-memory version chain is updated.

\begin{figure*}[t]
	\begin{center}
			\subfloat[\parbox{2.7in}{Index Performance under Mixed Workloads (CH-Bnchmark)}]{
			\includegraphics[width=0.9\columnwidth]{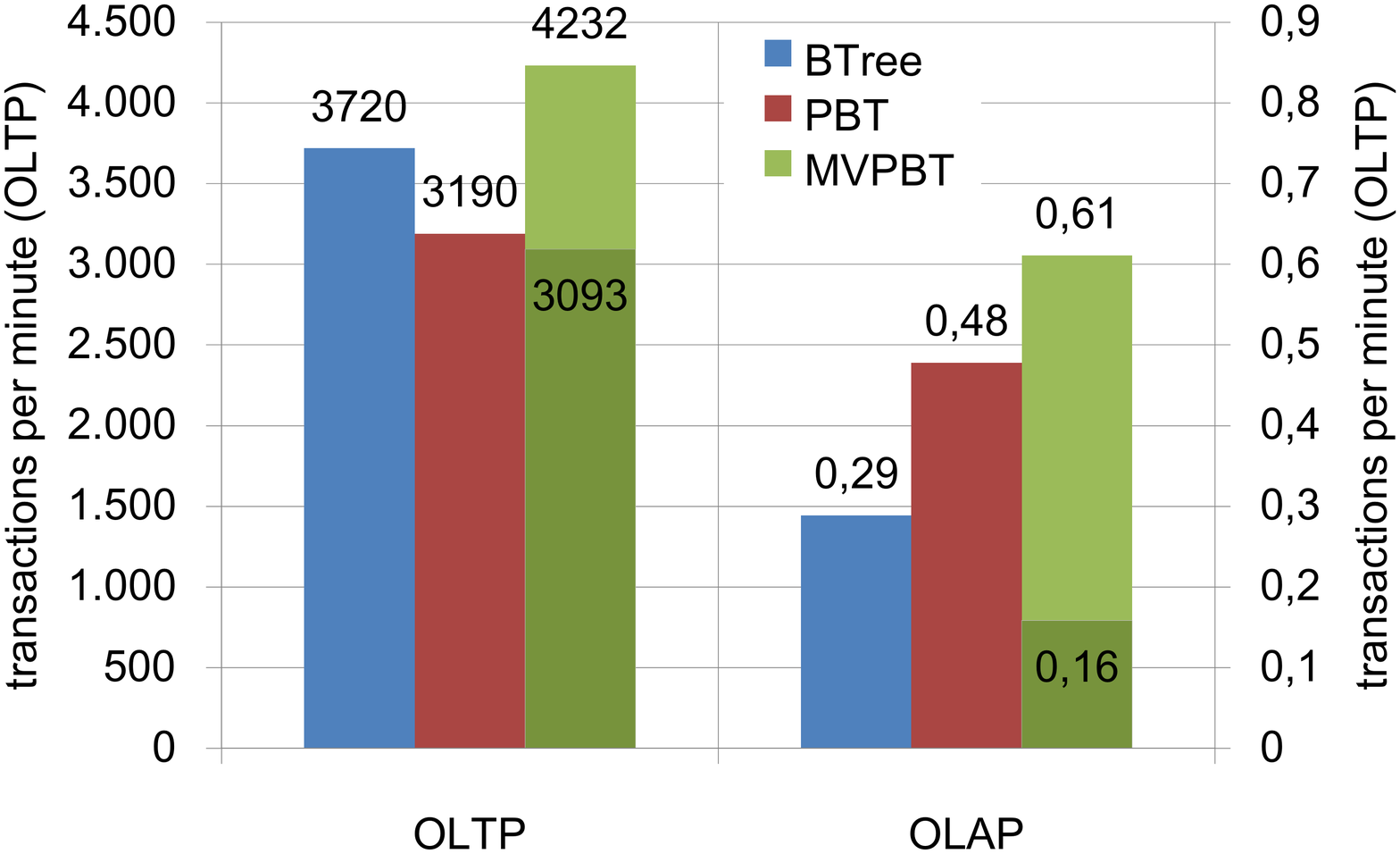}
			\label{img:test_ch}
		}
		\subfloat[\parbox{3in}{Standard vs. Index-Only Visibility-Check 
		for Different Chain Lengths}]{
			\includegraphics[width=0.9\columnwidth]{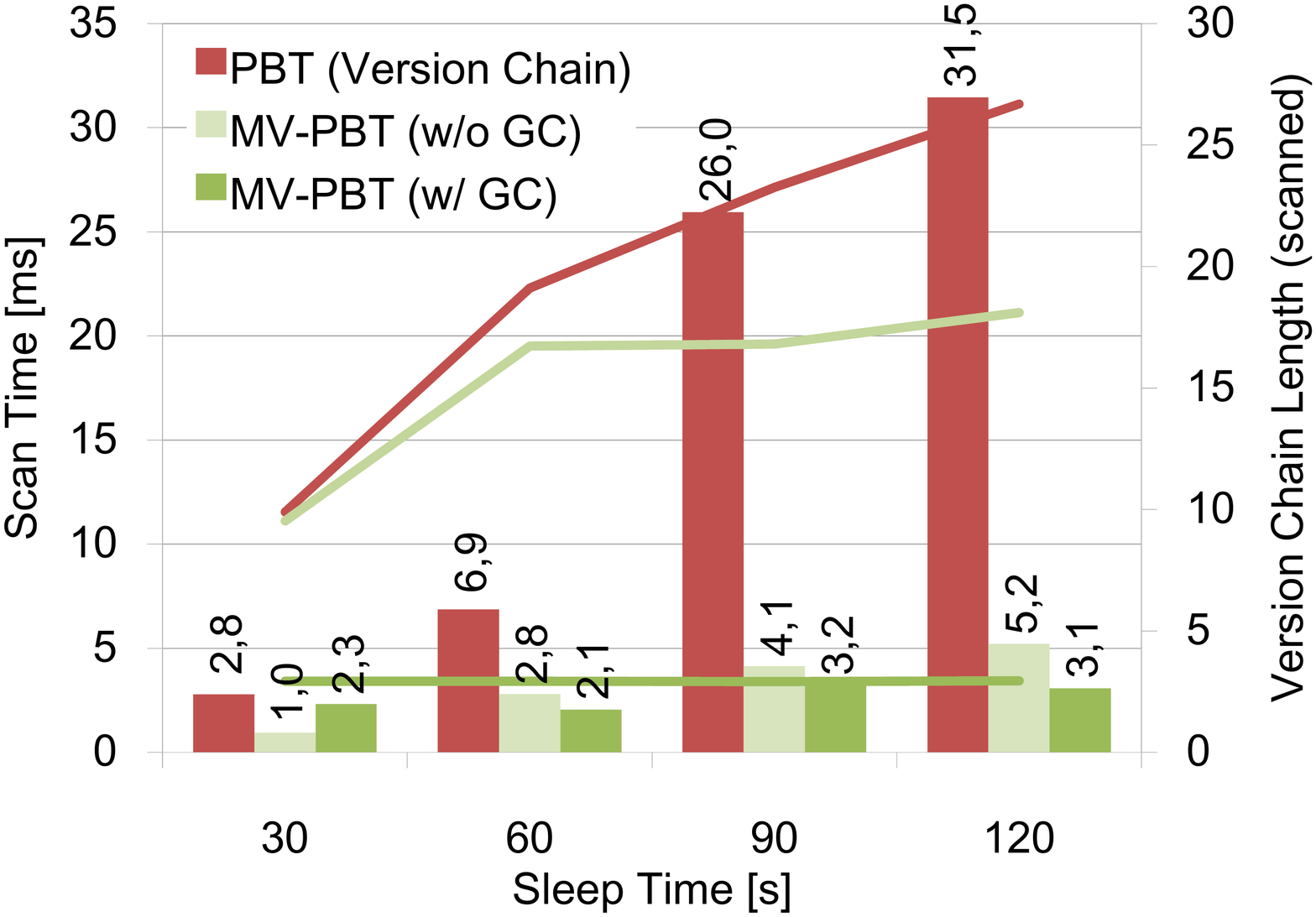}
			\label{img:test_vc_length}
		}
		\vspace{-10pt}
		\subfloat[\parbox{3in}{Sequential Write Pattern of Eviction of a Single MV-PBT Partition}]{
			\includegraphics[width=0.9\columnwidth]{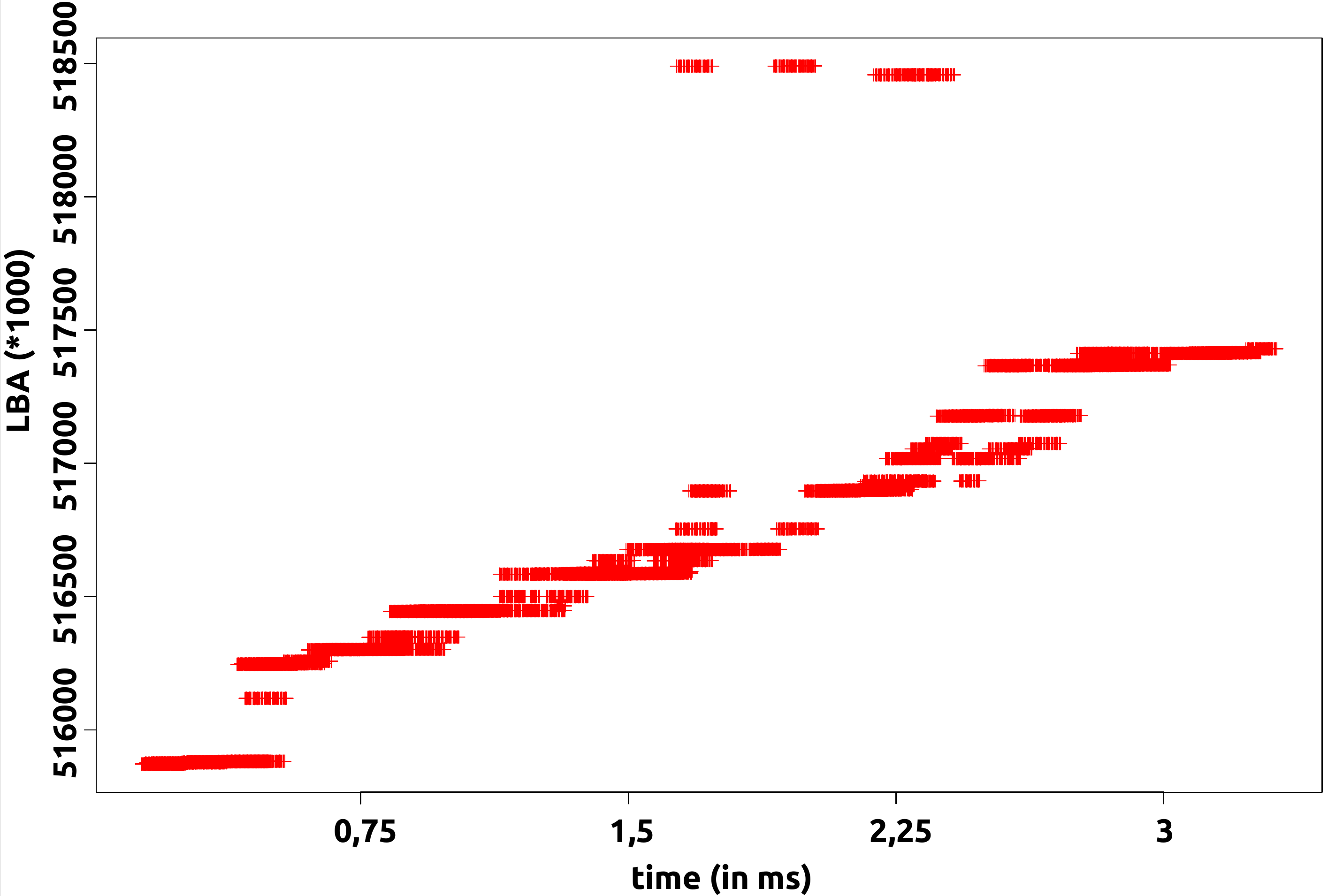}
			\label{img:test_write_io} 
		}
		\subfloat[\parbox{1.4in}{Requests / Cache Hit Rate}]{
			\includegraphics[width=0.9\columnwidth]{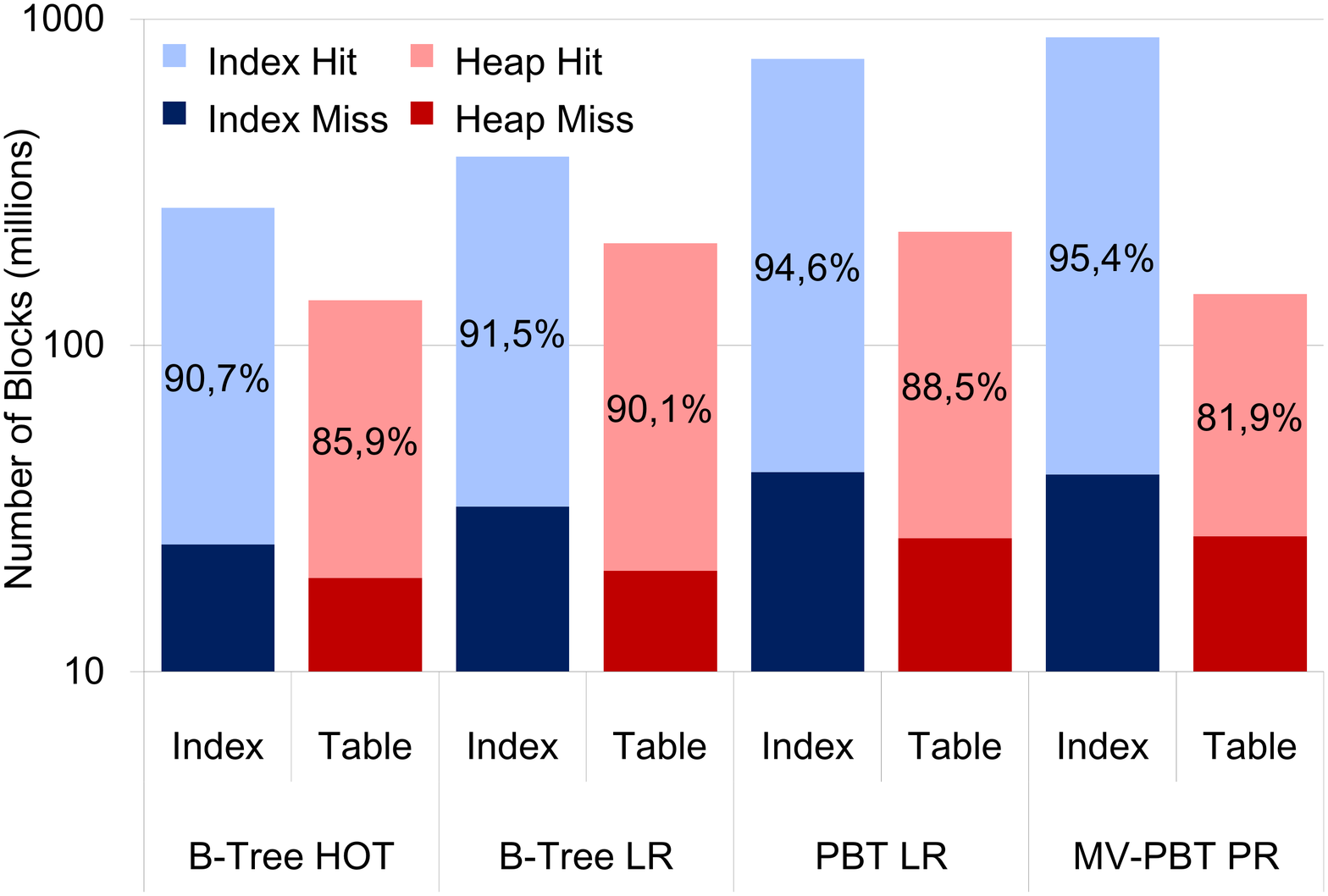}
			\label{img:test_cachehit} 
		}
		\caption{Index Performance under Mixed Workloads (CH-Benchmark)}
	\end{center}
\end{figure*}

\subsection{MV-PBT Filters and Optimizations}
\label{ssec:mvpbt_filters}
Various optimizations can be performed, based on the fact that once written to storage MV-PBT partitions are \emph{immutable}. 

{\bf Bloom Filters.} Each MV-PBT partition has a bloom filter (\emph{BF}) on the search key. Using bloom filters accelerates key lookups (point-queries) in a partition, by avoiding unnecessary scans. Whenever a key lookup is performed, a BF-query executed first, to verify whether the key does \emph{not} exist in the partition. If it does not exist MV-PBT proceeds with the next partition. Alternatively, if the BF returns true (i.e. the key may exist), MV-PBT scans the whole partition. 

Our experimental evaluation (Figure \ref{img:bf:effectiveness:size}) indicates that the average BF size is small -- in the order of few hundred KB. Therefore frequently used filters are usually cached in the MV-PBT buffer. Furthermore, their precision is \emph{98\%} on average, thus false positives and therefore superfluous scans are rare. BF is is computed efficiently on eviction, piggybacking existing maintenance scan and is persisted as part of the partition metadata. 

{\bf Range Filters.} Partition bloom filters accelerate point lookups, but cannot handle range predicates. Currently, we employ \emph{prefix Bloom Filters (pBF)}, if appropriate, to speedup range scans. 

{\bf Dense-packed, Read-Optimized immutable storage.} Since a partition is immutable once persisted, various space and read-optimization techniques can be applied. \emph{Dense-packing} is used to perform coalescing and free-space optimzation. When in-memory leaf nodes are on average 67\% full to accumulate modifications and avoid splitting, however when persisted the the space utilization can be maximized. MV-PBT performs \emph{dense-packing} as part of the final garbage collection and space reclamation.

Especially for \emph{non-unique indices} MV-PBT performs \emph{reconceliation} upon eviction to convert all regular records with the same search key to a single regular record with a set of \{recordID, timestamp\}, instead of holding separate record for each key instance. The same is true for replacement records, where for the same search key  sets of \{$recordID_{NEW}$, $Timestamp_{NEW}$, $recordID_{OLD}$\} are created. 
Last but not least, compression techniques such as \emph{prefix-truncation} or \emph{delta-compression} are performed on the search key. Along the same lines, various read  and cache-aware optimizations can be performed.

\section{Experimental Evaluation}
\label{sec:experimental_evaluation}

We present the analysis of Partitioned B-Trees (PBT) and Multi-Version Partitioned B-Trees (MV-PBT) together with traditional B\textsuperscript{+}-Trees in PostgreSQL 9.04, which serve as \emph{baseline}.  Standard, PostgreSQL uses an \emph{old-to-new} version ordering, \emph{physically materialized version storage} and \emph{two-point invalidation}. Index records have a physical reference to base tables -- denoted as B-Tree (PG/HOT). PostgreSQL base table storage was also modified to Snapshot Isolation Append Storage (SIAS) \cite{gottstein:siaschains,Gottstein:Diss:2016} with a beneficial append-only write pattern, one-point invalidation and new-to-old version ordering. We implemented and evaluated B\textsuperscript{+}-Trees and PBT with physical references and with logical tuple references on top of SIAS \cite{gottstein:siaschains,Gottstein:Diss:2016}.

{\bf Experimental Setup.} We deployed PostgreSQL 9.04 and PostgreSQL with SIAS \cite{gottstein:siaschains} on an \textit{Ubuntu 16.04.4 LTS} server with an eight core \textit{Intel(R) Xeon(R) E5-1620} CPU, 2GB RAM and an \textit{Intel 
DC P3600 400GB SSD} drive. We used the well-known DBT-2\cite{dbt2} TPC-C-like OLTP benchmark and mixed workload CH-Benchmark \cite{Cole::CH:DBtest:2011} in OLTP-Bench \cite{Difallah:oltpbnch:VLDB:2013,chbench} for experimental evaluation. The OS page cache was cleaned every second to ensure repeatable and reliable results (even though conservative).
 
{\bf Mixed Workloads: CH-Benchmark.} 
MV-PBT is designed for large datasets and mixed workloads. We evaluate the throughput of B\textsuperscript{+}-Trees, PBT and MV-PBT under the CH-Benchmark \cite{Cole::CH:DBtest:2011} in OLTP-Bench \cite{Difallah:oltpbnch:VLDB:2013,chbench}. MV-PBT \emph{doubles} the analytical throughput compared to B\textsuperscript{+}-Trees (Figure \ref{img:test_ch}), improving it from 0.29 to 0.61 queries/transactions per minute. In the same time, MV-PBT yield \emph{15\%} higher transactional throughput than B\textsuperscript{+}-Trees (Figure \ref{img:test_ch}).

The performance improvements are mainly due to \emph{index-only visibility-check} and \emph{partition garbage collection}. To illustrate the combined effect we turn off both and repeat the experiment. Consider now the lower MV-PBT performance bars in Figure \ref{img:test_ch}. Without \emph{partition garbage collection} and \emph{index-only visibility-check} the OLAP performance drops by 75\% from 0.61 to 0.16 queries per minute, whereas the OLTP throughput plummets from 4232 from to 3093 tx/min.

{\bf Mixed Workloads: Index-Only Visibility-Check and Garbage Collection.}
In a further experiment we investigate MV-PBT GC and visibility-check in more detail varying the version-chain length. We run the OLTP part of the CH-Benchmark and execute a query on the same dataset (Figure \ref{img:test_vc_length}), however we pause it (using \emph{pg\_sleep}) for 30/60/90/120 seconds to simulate a long-running query and vary the amount of transient versions and the chain length. Clearly, as the version-chain length increases, index-only visibility-checks gain importance, because unnecessary read I/O on base table can be reduced. 

We compare PBT and standard \emph{visibility-check in base table (VC)} to MV-PBT and \emph{index-only visibility-check (idxVC)} (Figure \ref{img:test_vc_length}). As the query processing time and version-chain length increase, index scans and \emph{VC} of  slow down PBT by an order of magnitude. Even if the version-chain length has no linear growth, pages in base table get evicted  and need to be fetched more frequently. MV-PBT performs \emph{idxVC} however without garbage collection (Figure \ref{img:test_vc_length} \emph{MV-PBT w/o GC}), every index record of successor tuple-versions has to be processed, likewise the scan time increases proportionally to the chain length. With garbage collection (Figure \ref{img:test_vc_length} \emph{MV-PBT w/ GC}), the number of scanned index records and the scan time remain almost constant. However, GC requires additional processing and latches index nodes in $P_{N}$. Reading transactions have to wait for latches and scan time increases - consider Figure \ref{img:test_vc_length} at 30 seconds sleep time. As more index record get garbage collected, GC improves the index scan time - compare MV-PBT with and without GC at 30 and 120 second (Figure \ref{img:test_vc_length}).

{\bf Sequential write-pattern/Append-based storage.} 
Based on the tradeoffs derived in Section \ref{sec:modern_hardware}. MV-PBT needs to support write sequentialization and append based storage.
In this experiment we evaluate the write pattern of MV-PBT (Figure \ref{img:test_write_io}). Using \emph{blktrace} and \emph{blkparse} we record an I/O trace during the partition eviction from MV-PBT buffer. The X-axis represents the eviction time; the average write I/O time is about 1ms. The Y-axis represents the logical block addresses (LBA), i.e. the file system addresses where the blocks of the index file are written. Each red cross indicates the write of a single index node. A horizontal line, therefore indicates a \emph{sequential write}, i.e multiple blocks are written onto neighbouring addresses over time. 
\emph{Hence the sequential write pattern of MV-PBT.} 
The horizontal lines in Figure \ref{img:test_write_io} represent \emph{database extents } and result from the database \emph{space allocation strategy}. Each evicted partition comprises leaf nodes allocated in new extents of the index file, allocated at (mostly) adjacent addresses by the file system. The overall \emph{sequential pattern} confirms the sequential append behaviour of MV-PBT.

{\bf MV-PBT Buffer Efficiency.} 
Figure \ref{img:test_cachehit} shows the fetch requests on index nodes (blue) and base table nodes (red) for an OLTP benchmark. Furthermore, the cache hit-rate is depicted. Requests yielding a cache-hit are displayed brighter colour than fetches (cache-misses) from secondary storage. The scale of requests is logarithmic. The results are calculated for equal throughput over the test duration and all tables and indices. 

PBT and MV-PBT require more requests on index nodes due to partitioning of index records and greater record sizes. Most requests are on buffered nodes, because many queries can be answered in the main memory partition. Index records of new tuple-versions are common to be located there. MV-PBT reduces the requests on base table by up to 40\%, because the base table is not required for visibility-check. The version chains are short for this benchmark, for mixed workloads this effect is more weighty. This can be seen at the reduced cache hit rate on base table nodes in comparison to PBT. Most saved requests on base tables are on new tuple-versions, which are located in main memory.

{\bf Partition Filters.} 
Partition-based indices like MV-PBT, PBT or LSM-Trees incur higher lookup and scan overhead than B-Trees, since matching records may exist in older partitions. Hence, the effort of lookups and especially of scans increases with number of index-partitions, since every partition has to be traversed in the worst case. Point lookups can stop partition traversal after finding the first matching record, which is visible to a transaction, since older partitions are guaranteed to contain older records. 

\begin{figure}[!h]
	\begin{center}
		\vspace{-10pt}
		\includegraphics[width=\columnwidth]{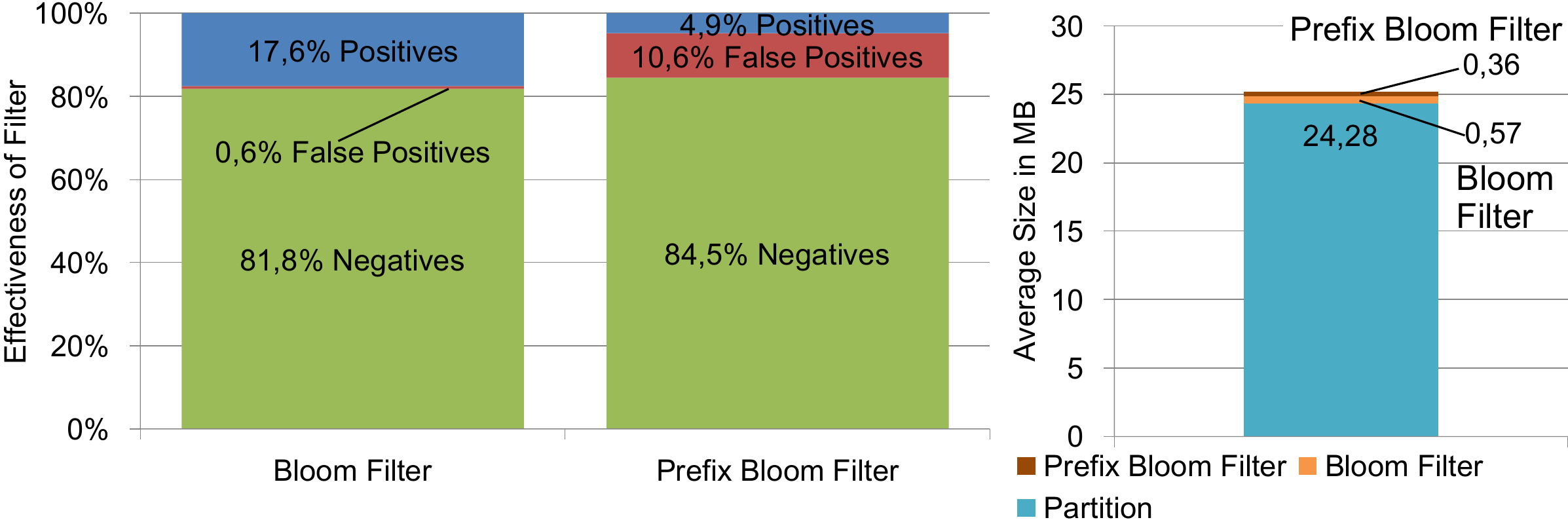}
		\vspace{-20pt}
		\caption{Effectiveness and Size of Partition Filters }
		\label{img:bf:effectiveness:size}
		\vspace{-10pt}
	\end{center}
\end{figure}

\begin{figure*}[!h]
	\begin{center}
		\subfloat[
		\vspace{-5pt}
		\parbox{2.5in}{Indirection Layer vs. Physical Version-Record Reference}]{
			\includegraphics[width=0.41\textwidth]{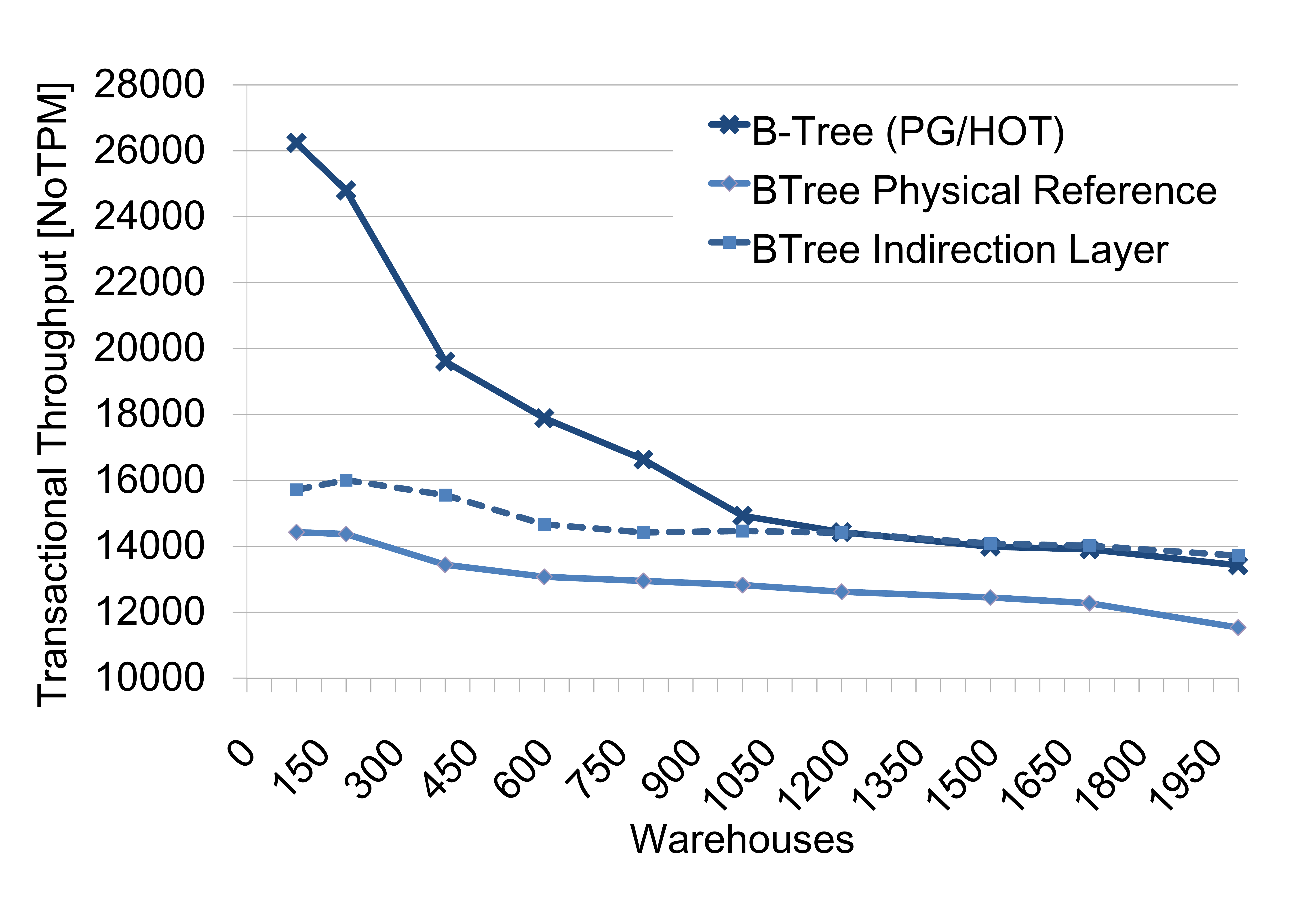}
			\label{img:test_indirection}
		}
		\subfloat[
		\vspace{-5pt}
		\parbox{2.3in}{Performance of Indexing Approaches under TPC-C}]{
			\includegraphics[width=0.41\textwidth]{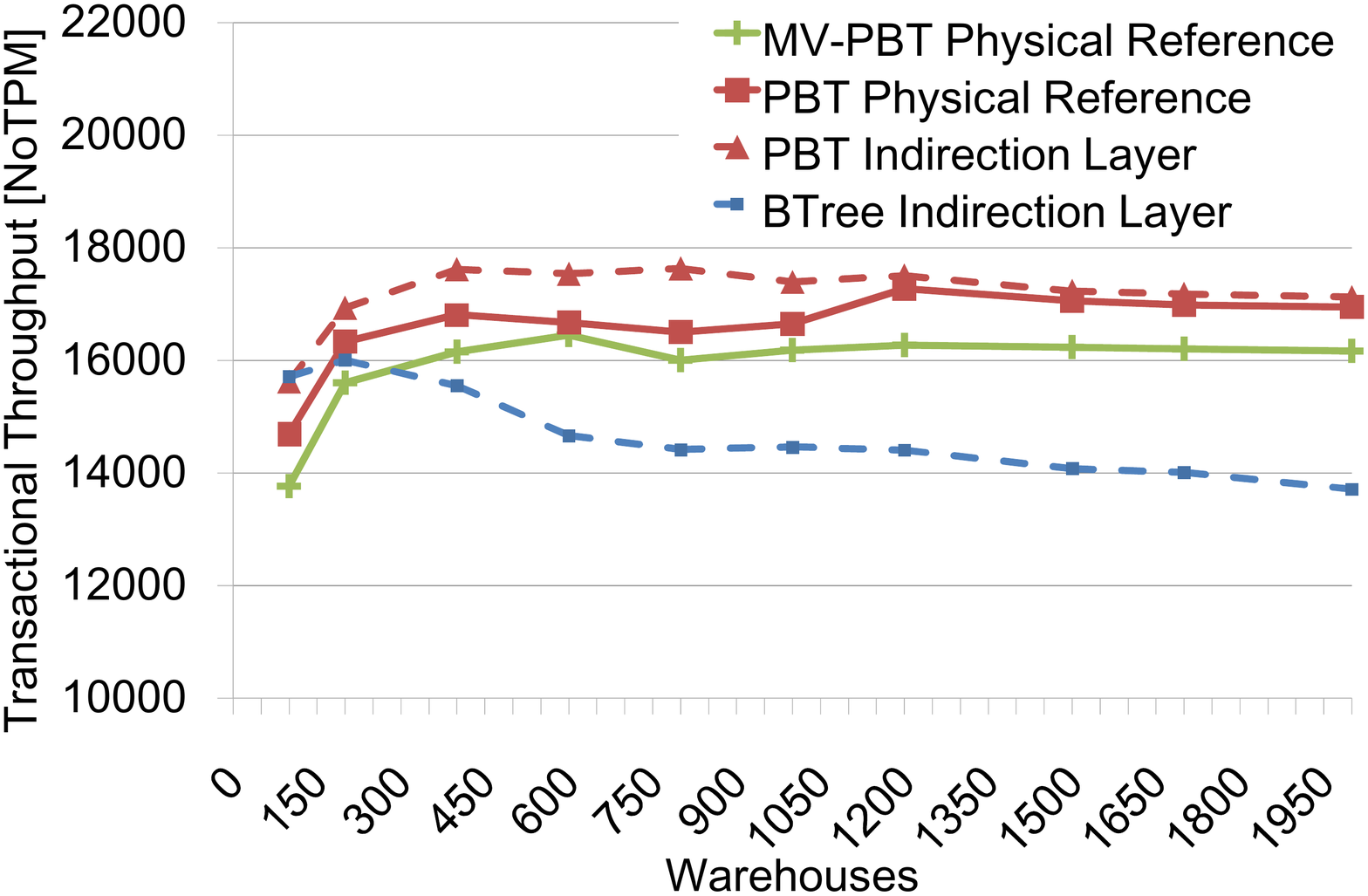}
			\label{img:test_indexing_apporaches}
		}
		
		\vspace{-7pt}
		\subfloat[ \parbox{2.6in}{Influence of filter techniques on Throughput under 
		TPC-C}]{
			\includegraphics[width=0.41\textwidth]{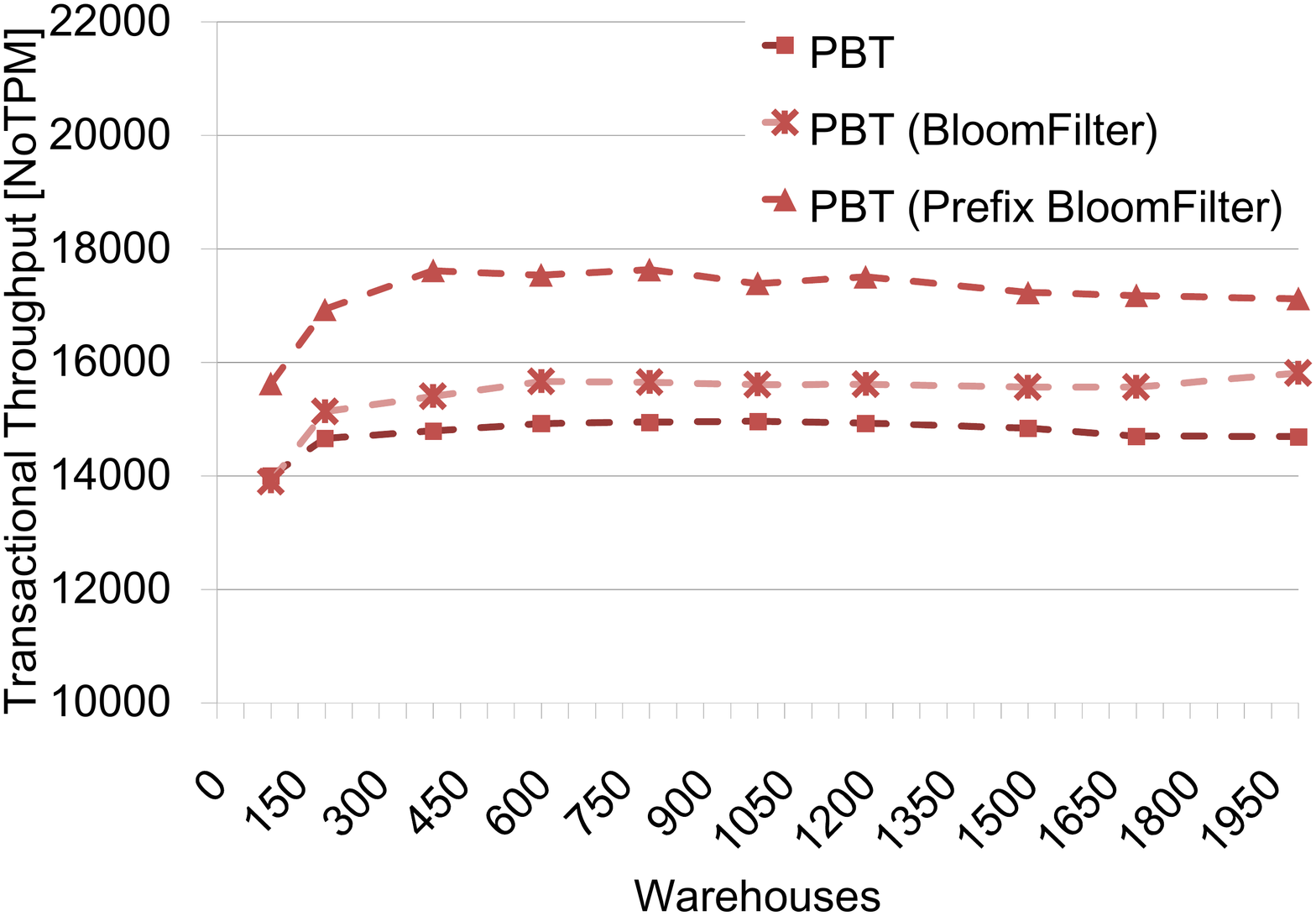}
			\label{img:test_filter}
		}	
		\subfloat[\parbox{1.85in}{MV-PBT Garbage Collection under TPC-C}]{
			\includegraphics[width=0.41\textwidth]{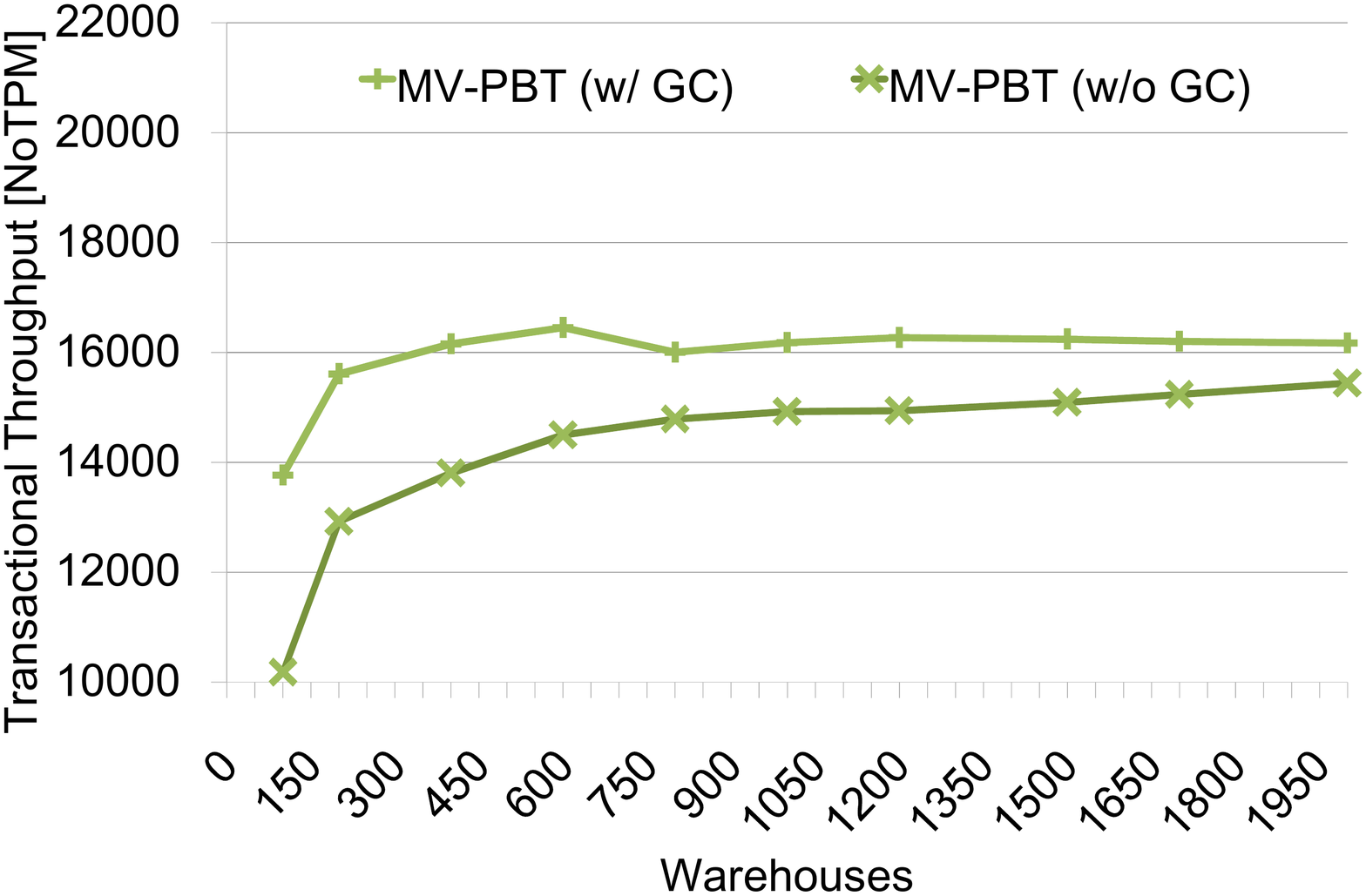}
			\label{img:test_mvpbtgc}
		}
		\caption{OLTP Performance Evaluation under TPC-C}
		\vspace{-7pt}
	\end{center}
\end{figure*}

Using \emph{Bloom filters (BF)} (Section \ref{ssec:mvpbt_filters}) point lookups can skip partitions and increase throughput up to 10\% under TPC-C (Figure \ref{img:test_filter}). Furthermore, \emph{prefix Bloom filters (pBF)} may under certain conditions speedup scans by skipping partitions not matching the range predicate. \emph{pBF} including a fixed set of scan attributes, increase the throughput by another 10\% (Figure \ref{img:test_filter}). The precision of both Bloom filters is relatively high (Figure \ref{img:bf:effectiveness:size}): the false positives rate is 2\% for \emph{BF}  and 10\% for \emph{pBF}, while the negatives (skipping) rate is approx. 82\% for \emph{BF} and 84.5\% for \emph{pBF}. The size \emph{BF} and \emph{pBF} is small relative to the partition size (Figure \ref{img:test_filter}): for a 24MB partition \emph{BF} is 0.57MB, while \emph{pBF} is 0.36MB.

Since index operations only have a fair share of the whole database operations under TPC-C (besides logging, CC and I/O) the above numbers yield moderate performance improvements.

\begin{figure}[!h]
	\begin{center}
		\subfloat[
		\vspace{-5pt}
		\parbox{1.7in}{MV-PBT, BTree, LSM-Tree under YCSB.}]{
			\includegraphics[scale=0.3]{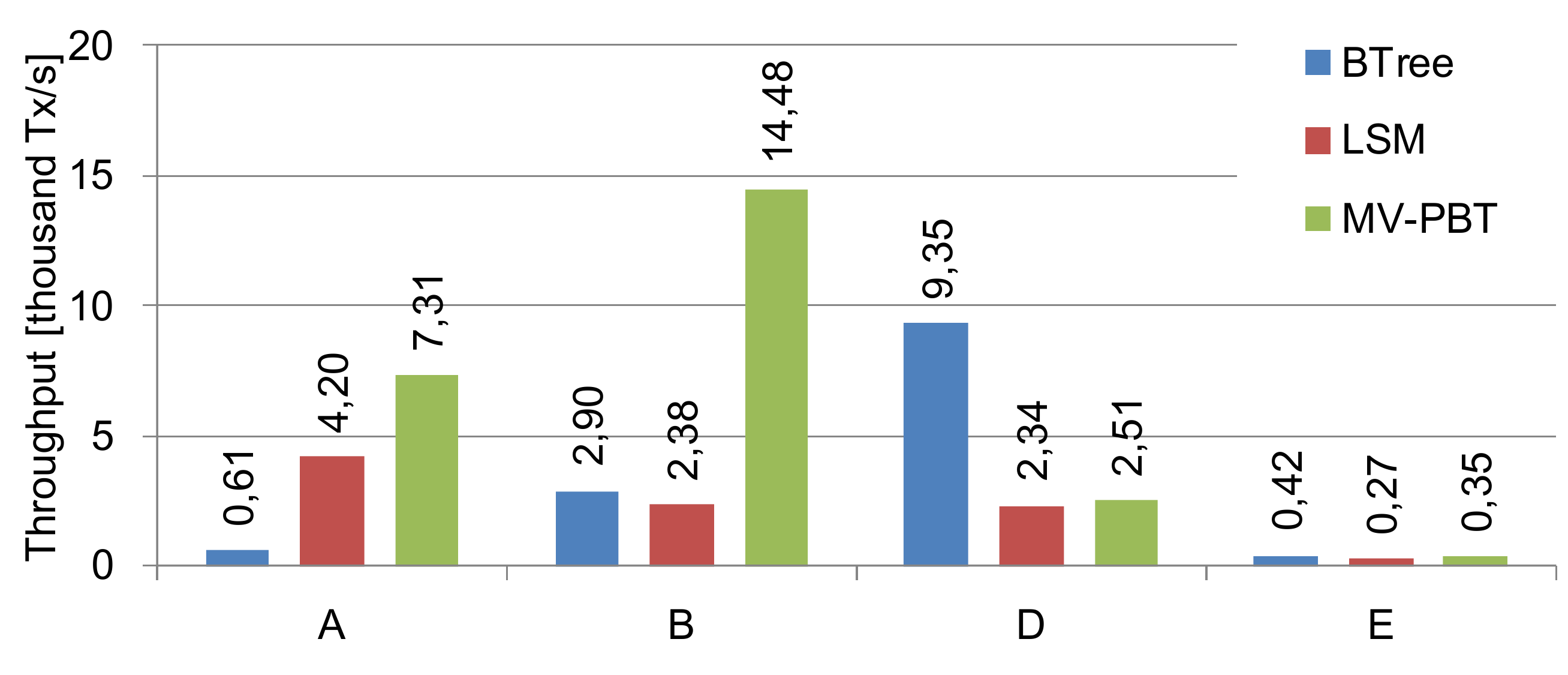}
			\label{fig:lsm}
		}
		
		\vspace{-3pt}
		\subfloat[ \parbox{2.9in}{YCSB Throughput (workload A) vs. Number of MV-PBT Partitions}]{
			\includegraphics[width=0.45\textwidth]{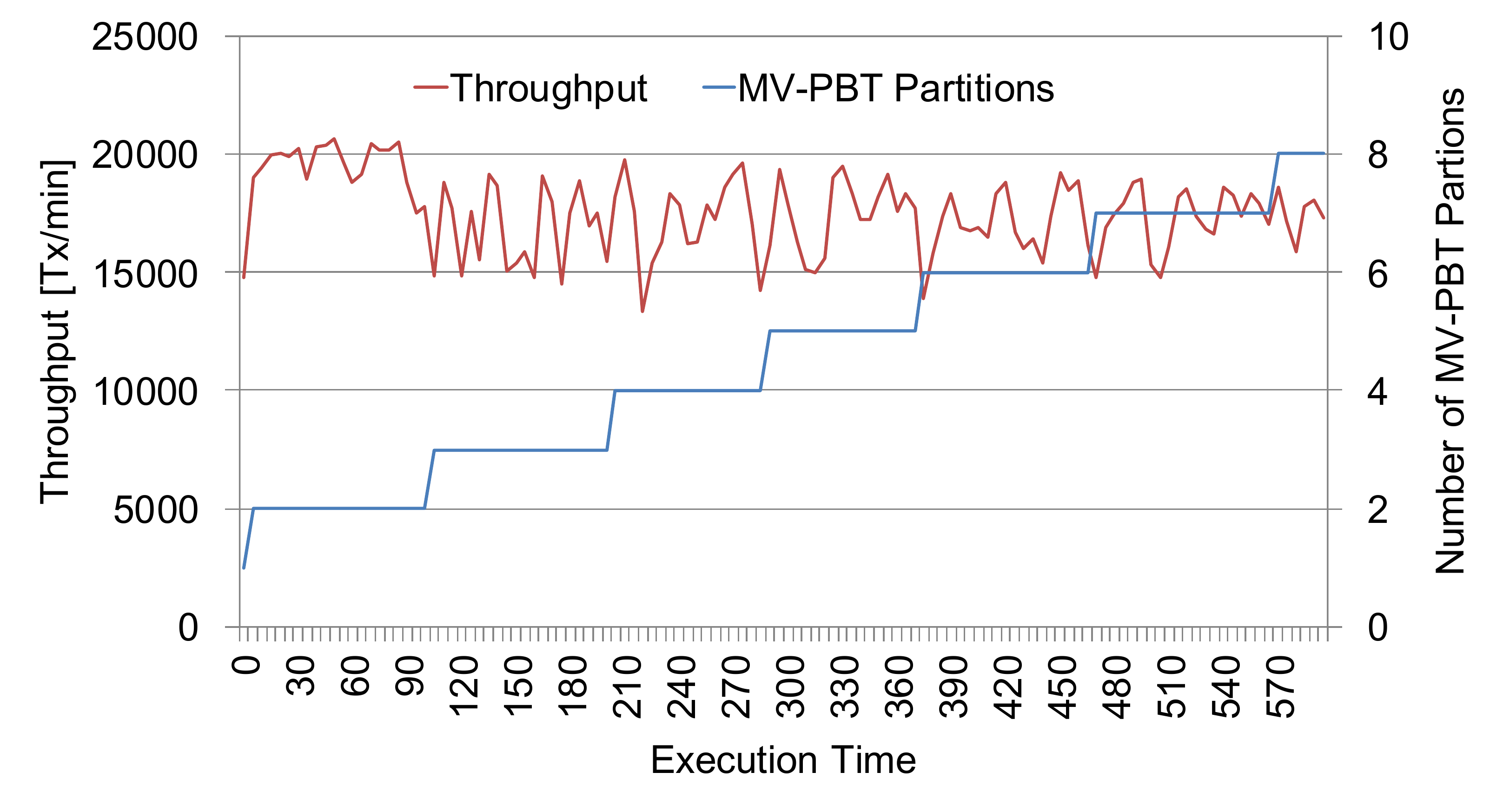}
			\label{fig:partitions}
		}	
		\caption{Performance Evaluation under YCSB}
		\vspace{-10pt}
	\end{center}
\end{figure}

{\bf Comparison to LSM-Trees.} 
LSM-Trees \cite{Luo2019} are used as workhorse storage structure in many Key/Value stores for large datasets. Today's highly-optimized multi-level LSM-Trees with levelling or tiering resemble MV-PBT as they exhibit an append-behaviour and employ buffered components. 
We implemented MV-PBT in WiredTiger \cite{wiredtiger}, the high-performance KV-Store of MongoDB. In this experiment we compare MV-PBT to LSM-Tree in WiredTiger under YCSB (Figure \ref{fig:lsm}). YCSB has been instrumented as follows: a dataset of 100 million keys (approx. 100GB); workloads A (30 {mil.} requests), B and D(10 {mil.} req.) and E (2 {mil.} req).


Workload \emph{A} comprises 50\% read and 50\% update requests, which require fast lookups and updates. MV-PBT is approx. 42\% faster than LSM-Trees. Each LSM level comprises multiple components which themselves are small read-optimzed BTrees.  A search needs to process separate LSM components even though some can be skipped (bloom filters). 
MV-PBT partition search is faster than LSM component search since the leaf nodes in each partition are under the same common index. Updates in MV-PBT, hit $P_{N}$, which accommodates more kv-pairs than the main memory $L_{0}$ in LSM-Trees. 
Workload \emph{B} comprises 95\% read and 5\% update requests, with \emph{zipfian} distribution. BTree performs random reads, the LSM Tree caches the updates but has an equal amount of random reads spread over more componets. MV-PBT have much lower index maintenance compared to BTrees and place the updates in $P_{N}$. The reads are performed with maximum I/O parallelism.
Workload \emph{D} comprising 95\% read and 5\% update requests, which given the \emph{latest} distribution stress the memory components and BTrees performs most of the operations in memory. MV-PBT is marginally better than LSM-Tree. Last but not least, we run workload \emph{E} comprising 95\% scans and 5\% insert requests. Even though the \emph{scans} are slow under MV-PBT, they outperform LSM-Trees due to the faster search and updates.

Consider, Figure \ref{fig:partitions} depicting the YCSB throughput (workload A) and the number of MV-PBT partitions over time. The throughput remains stable as the number of partitions increases. 

{\bf OLTP: comparison of B-Tree alternatives.} To establish the baseline we first compare standard PostgreSQL B-Trees (PG/HOT) to B\textsuperscript{+}-Trees with physical reference and indirection layer on top of append-only storage (SIAS \cite{gottstein:siaschains,Gottstein:Diss:2016}) under TPC-C. In Figure \ref{img:test_indirection}, we show the throughput for different dataset sizes. The buffer cache of the DBMS is fixed to 600MB. B-Tree(PG/HOT) performs well (Figure \ref{img:test_indirection}) as long as the database buffer can accommodate most modifications. Under standard Postgres updates are performed in base tables by Heap-Only Tuples, i.e. the predecessor version is cached on the same page where its successor is located. Therefore the index maintenance effort is low. With growing data sizes (and therefore more modifications), the throughput falls rapidly. Append-based storage and one-point invalidation (SIAS \cite{gottstein:siaschains,Gottstein:Diss:2016}) exhibit a \emph{robust} throughput: (a) \emph{physical references} (Section \ref{ssec:index_management}) yield lower performance, due to the higher index management overhead; (b) an \emph{indirection layer} reduces index maintenance for insertions and index-key updates, yielding up to 30\% better throughput. \emph{With larger datasets ($\geq$ 1200 warehouses) B-Trees with indirection outperform standard PostgreSQL PG/HOT.}

{\bf Indexing Approaches under OLTP.} In a follow-up experiment, we compare B-Tree with indirection layer (Section \ref{ssec:index_management}), to \emph{PBT} and \emph{MV-PBT} under TPC-C (Figure \ref{img:test_indexing_apporaches}). PBT and MV-PBT exhibit robust performance, which improves with larger datasets compared to B-Tree.
PBT with \emph{indirection layer} exhibits high and robust performance (Figure \ref{img:test_indexing_apporaches}). PBT with \emph{physical reference} to close the performance gap for larger datasets as the update density decreases decreases with larger datasets.
\emph{MV-PBT are slower than PBT under OLTP workloads for several reasons}. First, less MV-PBT index records fit on the same sized $P_{N}$, since their sizes are larger because of the version-information (transaction timestamps). Consequently, the number of partitions increases, yielding more I/O. Second, the average version-chain length under TPC-C is short: 1.15/2.18 versions for \emph{customer}/\emph{stock} respectively \cite{Gottstein:Diss:2016}. Therefore, index-only visibility-checks cannot improve performance significantly. Thus, MV-PBT exhibit 6\% lower performance than PBT under TPC-C (Figure \ref{img:test_indexing_apporaches}). 
We implemented MV-PBT with an \emph{indirection layer} as well as with \emph{physical references} (Section \ref{ssec:index_management}). Figure  \ref{img:test_indexing_apporaches} depicts on the performance with \emph{physical references} for brevity, both curves are almost identical. \emph{Therefore, MV-PBT are general enough to be implemented matching the rest of the system design.}

 {\bf OLTP Garbage Collection.} 
In this experiment (Figure \ref{img:test_mvpbtgc}) we quantify the performance effect of MV-PBT partition garbage collection (Section \ref{ssec:mvpbt_in_memory_gc}). It improves performance between 5\% and 17\% since old invisible versions are purged and need not be processed by scans as well as space is reclaimed letting more index records fit in $P_{N}$. The opportunity of improvement under OLTP is however limited by the short average version-chain length: 1.15 versions for \emph{customer} and 2.18 versions for \emph{stock} under TPC-C \cite{Gottstein:Diss:2016}. With HTAP workloads the amount of 'transient' (short-lived versions visible only throughout the duration of an analytical query) versions increases rapidly as does the effect of garbage collection. Garbage collecting larger amounts transient versions has a major role on the performace improvment of MV-PBT over PBT and B-Tree under mixed workloads (Figure \ref{img:test_ch}).

\section{Related Work}
\label{sec:related_work}
Most popular indexing approaches in database management systems (DBMS) are based on B\textsuperscript{+}-Trees. Their alphanumeric sorted structure can result in high write amplification for high update rates and visibility-checks require information, that is only located at tuple-versions in base table. PostgreSQL uses Heap-Only Tuples (HOT) to reduce index management operations. Index records reference items in base table, which point to tuple-versions in the heap node. Corresponding tuple-versions are held on the same node and can be located by processing the version chain. If a tuple-version become garbage collected, the item is modified to reference the next version. This indirection layer reduces index modifications, but cannot avoid write amplification of index nodes and requires the base table for visibility-checking. Furthermore the write amplification of base table nodes is increased for large datasets. MV-IDX\cite{gottstein:mv} maintains a virtual identifier for each tuple and data nodes for each version as an indirection layer. With Snapshot Isolation Append Storage (SIAS)\cite{gottstein:siaschains} write amplification on base tables is reduced in comparison to HOT, but index management operations can cause a high write amplification and base table nodes are still required for visibility-checking\cite{crizz:survey_mv_idx}. LSM-Trees\cite{oneil:lsm} reduce write amplification due to collecting modifications in main memory components, but there is no concept for managing tuple-versions and perform an index-only visibility-check\cite{crizz:survey_mv_idx}. Time-Split B-Trees \cite{lomet:timesplit} and Multiversion B-Trees \cite{seeger:mvbt} are able to separate index records of old tuple-versions from current dataset and to perform an index-only visibility-check, but maintenance operations are complex and can cause a high write amplification of index nodes\cite{crizz:survey_mv_idx}.

\section{Conclusion}
\label{sec:conclusion}
In the present paper we introduce MV-PBT as an approach to multi-version indexing. A MV-PBT is an extension of a B-Tree, where an artificial leading column is prepended to the search key of each index record and index records a placed in a buffered index partition, which if full gets evicted and appended to persistent storage. MV-PBT is version-aware, since index records contain version-information and allow for index-only visibility check. This is particularly beneficial for HTAP workloads since long chains of transient versions exist due to the mix of short-lived updating transactions and long-running queries. Furthermore, MV-PBT exhibit a sequential write pattern due to the concept of partition, which leads to less write-amplification and better utilization of modern storage technologies. Under mixed workloads (CH-Benchmark) MV-PBT double the analytical throughput 2x, while improving the transactional throughput by 15\%.



\bibliographystyle{ACM-Reference-Format}
\bibliography{references}

%
% ...

\end{document}